\documentclass[twocolumn]{svjour3}

\usepackage{graphics} 
\usepackage{epsfig} 
\usepackage{amsmath} 
\usepackage{amssymb}  
\usepackage{gensymb}
\usepackage{subfig}
\usepackage[export]{adjustbox}
\usepackage{paralist}
\usepackage{commath}

\usepackage{color}
\usepackage{booktabs}
\usepackage{flushend}
\usepackage{multirow}
\usepackage{enumitem}
\usepackage{longtable}
\usepackage{afterpage}
\usepackage[T1]{fontenc}
\usepackage{natbib}
\setcitestyle{square}
\usepackage{multicol}
\usepackage{xltabular}
\usepackage{pdflscape}
\usepackage[counterclockwise]{rotating}

\renewcommand{\qed}{\hfill\ensuremath{\blacksquare}}
\usepackage{longtable}
\usepackage{color}

\journalname{Journal of Cryptographic Engineering}

\newcommand{\Keywords}[1]{\par\noindent{\small{\bf Keywords\/}: #1}}

\newcommand{\origttfamily}{}
\let\origttfamily=\ttfamily
\renewcommand{\ttfamily}{\origttfamily \hyphenchar\font=`\-}

\graphicspath{{Figs/}}
\DeclareGraphicsExtensions{.pdf,.png,.jpg,.eps}

\begin{document}

\title{Physical Security in the Post-quantum Era}
\subtitle{A Survey on Side-channel Analysis, Random Number Generators, and \\Physically Unclonable Functions}

\author{Sreeja Chowdhury$^*$ \and Ana Covic$^*$ \and Rabin Yu Acharya \and Spencer Dupee \\ Fatemeh Ganji$^\dag$ \and Domenic Forte}

\institute{ 
S. Chowdhury,  A. Covic, R. Yu Acharya, S. Dupee, and D. Forte \at 
$^*$ These authors have equally contributed to this work. \\
Florida Institute for Cybersecurity Research,\\
University of Florida \\
601 Gale Lemerand Dr, Gainesville, FL 32603\\
USA\\
\email{\{sreejachowdhury, anaswim, rabin.acharya,\\spdupee, dforte\}@ufl.edu}\\
\and 
F. Ganji \at 
$^\dag$ Corresponding author\\
Electrical and Computer Engineering, \\ 
Worcester Polytechnic Institute\\
100 Institute Road, Worcester, MA 01609-2280\\
USA\\
\email{fganji@wpi.edu}
}

\maketitle
\begin{abstract}
Over the past decades, quantum technology has seen consistent progress, with notable recent developments in the field of quantum computers. 
Traditionally, this trend has been primarily seen as a serious risk for cryptography; however, a positive aspect of quantum technology should also be stressed. 
In this regard, viewing this technology as a resource for honest parties rather than adversaries, it may enhance not only the security, but also the performance of specific cryptographic schemes. 
While considerable effort has been devoted to the design of quantum-resistant and quantum-enhanced schemes, little effort has been made to understanding their \emph{physical} security. 
Physical security deals with the design and implementation of security measures fulfilling the practical requirements of cryptographic primitives, which are equally essential for classic and quantum ones. 
This survey aims to draw greater attention to the importance of physical security, with a focus on secure key generation and storage as well as secure execution. 
More specifically, the possibility of performing side-channel analysis in the quantum world is discussed and compared to attacks launched in the classic world. 
Besides, proposals for quantum random number generation and quantum physically unclonable functions are compared to their classic counterparts and further analyzed to give a better understanding of their features, advantages, and shortcomings. 
Finally, seen from these three perspectives, this survey provides an outlook for future research in this direction\footnote{
This is a post-peer-review, pre-copyedit version of an article published in Journal of Cryptographic Engineering. The final authenticated version will be available online at: https://doi.org/10.1007/s13389-021-00255-w. 

\textsuperscript{\textcopyright}2021 Springer-Verlag GmbH. Personal use of this material is permitted. Permission from Springer-Verlag GmbH must be obtained for all other uses, in any current or future media, including reprinting/republishing this material for advertising or promotional purposes, creating new collective works, for resale or redistribution to servers or lists, or reuse of any copyrighted component of this work in other works.}. 
\vspace{5pt}
\end{abstract}

\section{Introduction}\label{sec:intro}
\sloppypar{
The omnipresence of computers, computing platforms, and services has been shaping the way that we handle various tasks ranging from simple switching to weather forecasting performed by controllers in modern appliances and supercomputers, respectively. 
For the latter purposes, it is widely accepted that several computational problems are fundamentally hard, even for modern machines with state-of-the-art computational power and resources. 
With the development of quantum physics, it is claimed that further computational power can be gained through describing the behavior of systems at a higher level of granularity (i.e., atomic and subatomic levels), where classical physics fails~\citep{maslov2018outlook}. 
Since the introduction of this idea, it has been doubted whether a practically feasible computational machine operating on quantum principles can be built~\citep{ladd2010quantum,dyakonov2019will}. 
In spite of all criticisms, the efforts to realize quantum computers have been made progressively, but an admittedly considerable breakthrough was recently reported in~\citep{arute2019quantum,moore2019intel}. 
Interestingly enough, it has been demonstrated that their proposed processor, named ``Sycamore'', comprised of fast, high-fidelity quantum logic gates can perform a computational task in 200 seconds, which would take the world's fastest supercomputer 10,000 years to finish. 

Perhaps the most drastic consequence of this technological advancement is that now, one of the main challenges in the realization of attacks on some of the cryptographical schemes could be dealt with. 
More specifically, for more than two decades, it has been thought that the quantum computers required to run Grover's and Shor's algorithms - which can break the security of symmetric and public-key cryptography, respectively~\citep{grover1996fast,shor1999polynomial} - could not be practically achievable~\citep{schneier2018cryptography}. 
The Sycamore processor can be seen, of course, as a first step towards building powerful quantum computers, which can compromise the security of cryptographic protocols applied in our every-day life. 
Besides these protocols, other crucial questions to ask would be: in the face of quantum-enabled attacks, which physical primitives remain secure. 
In fact, while the impact of the quantum computing paradigm on various cryptographic protocols and primitives has been intensely studied, its effect on physical security is less well understood. 
Physical security concerns developing measures to meet the needs of cryptographic primitives in practice. 
More concretely, such measures should be in place to narrow the gap between the characteristics of a cryptographic primitive implemented in reality and what is \emph{assumed} about that cf.~\citep{maes2013physically}. 
For these physical measures, referred to as ``Root-of-Trust'' (ROT), in the face of attacks becoming feasible in the post-quantum era, the physical-security assessment should be revisited. 
Particularly, the following objectives are essential for such an evaluation (see Figure~\ref{fig:rot}). 

\vspace{0.5ex}\noindent\textbf{Secure key generation:} True random number generators (TRNGs) are one of the most well-acknowledged and promising candidates introduced to harvest random numbers from physical sources of randomness. Albeit being trusted to generate random numbers with high quality (i.e., being uniformly distributed), there are difficulties in implementing such generators. One is how one should examine whether the generator exhibits the desired uniformity, which has been addressed by introducing the notion of ``adversarially controlled sources of randomness'' in the literature. In the classic world, this issue has been well studied; however, in a quantum world, more effort must be put into tackling this problem. More precisely, we are interested in procedures, which can guarantee that the bits produced by the generator are close to being indistinguishable from uniform bits from the point of view of a quantum adversary. 

\vspace{0.5ex}\noindent\textbf{Secure key generation and storage using physically unclonable functions (PUFs):} The premise underlying the concept of PUFs is that they can generate unpredictable and instance-specific random numbers to offer physically secure key generation and storage. In the classic world, after the introduction of the first PUF, it has become evident that PUFs are vulnerable to a wide range of attacks covering physical, invasive attacks to non-invasive machine learning (ML) attacks. Several countermeasures, from structural to protocol level, have been proposed to increase the security of PUFs against various types of attacks. Nevertheless, when assessing the security of these proposals, the main question to ask is that if they could remain secure in a quantum world. In this regard, not only should new PUFs be designed and implemented, but also threat models and risk assessments of possible attacks have to be considered carefully. 

\vspace{0.5ex}\noindent\textbf{Secure execution:} Nowadays, it is widely accepted that in the classic world, the general assumption about the secure execution of cryptographic implementations has been refuted by mounting numerous attacks. This has initiated a new line of research with the aim of enhancing the design of circuits to minimize side-channel leakages. This direction should be further pursued to investigate how mathematical algorithms employed in the side-channel analysis, and in particular, their computational complexity would evolve in a quantum world. This is indeed a crucial step to provide a better understanding of the attacker's capabilities in the post-quantum era. 
\begin{figure}[t]
    \centering
    \includegraphics[width=\columnwidth]{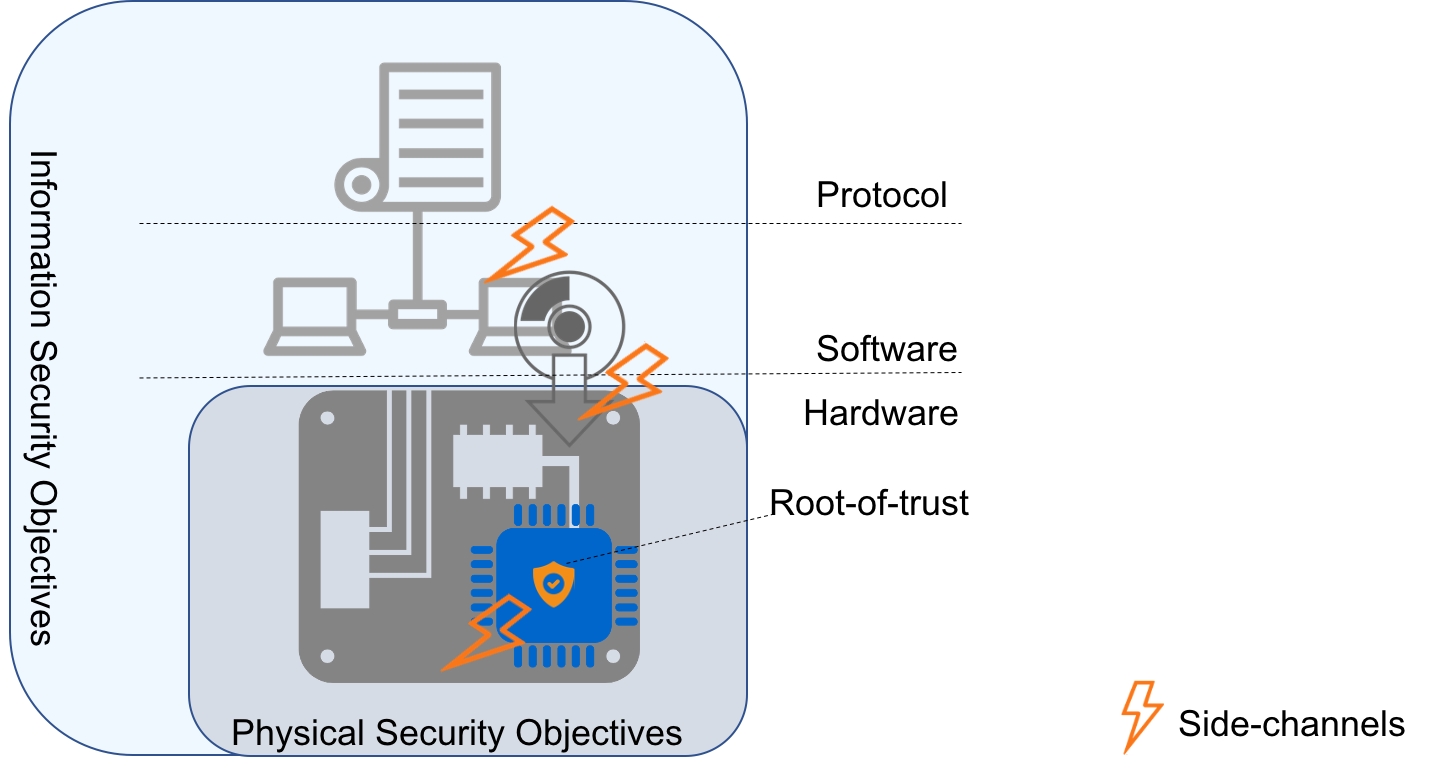}
    \caption{The tree-of-trust built on a root-of-trust (RoT). 
    RoT plays an important role in fulfilling the objectives of physical security. 
    These objectives include secure key generation and storage as well as secure execution, i.e., (ideally) eliminating certain physical side-channels~\citep{maes2013physically}. 
    Some examples of RoTs are PUFs, TRNGs, and side-channel-resistant schemes. 
    Note that side-channels can be leaked at different levels of the tree-of-trust, e.g., interpreter-level side-channels and timing side-channel leaked from a client and an algorithm in software, respectively~\citep{verbauwhede2007design}. 
Nevertheless, in this survey, we are interested in physical side-channels.  \vspace{-20pt}    }
    \label{fig:rot}
\end{figure}

So far, we mainly emphasize how an adversary equipped with quantum technology can compromise the security of physical primitives and RoTs. 
It is equally important to understand how this technology can enhance security by accomplishing tasks that cannot be performed in the classic world. 
In this case, quantum computers are not our main focus, but quantum devices, in particular, RoTs that take advantage of enhanced properties offered by quantum physics. 
In other words, providing the same functionality as their classical counterparts, quantum-enhanced RoTs should exhibit better features, e.g., security or efficiency. 
One popular example of such RoTs is quantum random number generation through key expansion, made possible in the quantum world (see, Section~\ref{sec:trngs} for more details). 
The implementation of quantum-enhanced RoTs gains pace thanks to various innovative products that should rely on the security of these primitives. 

\vspace{0.5ex}

\noindent\textbf{A brief overview and the organization of the paper:} By drawing attention to the positive side and negative side of the quantum technology for physical security, this survey investigates the state-of-the-art techniques proposed in the literature. 
Before giving an overview of the content of each section, we stress that the goal of our survey covers neither an inclusive list of all studies in quantum cryptosystem nor the detailed design of systems developed to enhance the physical security in both of the classic and quantum world. 
Yet, our survey aims at examining the research landscape in physical security and strengthening research in this direction. 
In this regard, it forms the basis for a systematic and comprehensive study on specific aspects of physical security in the quantum era, namely secure key generation and storage as well as secure execution. 
For excellent surveys covering a broader spectrum of research areas related to cyber-security and quantum cryptosystems, we refer the reader to~\citep{mosca2018cybersecurity,wallden2019cyber,nejatollahi2019post}, just to name a few. 

In our survey, after discussing how cryptography has been leveraging the advantages of quantum technology in Section~\ref{sec:qcrypto}, Sections~\ref{sec:pufs}--\ref{sec:sca} describes the recent developments in physical security and RoTs. 
In each of these sections, a taxonomy is proposed, which reflects the nature of existing methods in terms of how quantum technology enables us to either enhance or assess the security. 
In the latter case, we primarily look into attacks that can be launched thanks to the progress made in quantum technology. 
Finally, Section~\ref{sec:conclusion} expands on lessons learned and future directions in this area of research. 

}

\section{Quantum Computing for Cryptography: A Brief Overview}\label{sec:qcrypto}
\sloppypar{
As briefly discussed in Section~\ref{sec:intro}, even carefully-chosen cryptosystems devised for day-to-day applications in the classic world could be susceptible to attacks, which become feasible in the quantum world.  
These attacks can be seen as immediate consequences of the development of quantum computers with more computational power than their classical counterparts. 
The advantages of quantum technology are, however, not limited to this since it can be applied to equip honest parties\footnote{In this survey, as usually mentioned in cryptography-related literature, we refer to users and attackers as honest and malicious parties, respectively. } with quantum-enhance devices to obtain significant improvement in comparison to a classical setting. 
This section deals with the effects that quantum technologies have on the design of such devices as well as attacks against cryptosystems.

\subsection{Quantum Computing against Classical Schemes}
The practicality of our most common cryptographic schemes, i.e., RSA, Diffie-Hellman, and Elliptic curve cryptography (ECC), relies on the difficulty of two problems: the factoring problem and the discrete logarithm problem~\citep{ref1}. 
While factoring and discrete logarithms are not in themselves interesting problems, they have been found to be crucial for public-key cryptography. This application, in turn, remains sufficiently secure as far as the mathematical problems underlying their design remain difficult. 
In other words, the computational-power requirements of established cryptographic algorithms prevent attackers from stealing our data and allowing the security of public-key systems and privacy of transactions for all.
However, it is known that the advent of certain quantum algorithms has theoretically transformed the exponential time complexity of these cryptographic schemes; hence, they cannot be assumed secure in the quantum world. 

Notable quantum algorithms, which cause these security concerns, are Shor's algorithm and Grover's algorithm~\citep{shor1999polynomial,grover1996fast}.  
Shor found a clever way to factor numbers in $O(({\log{N}})^{3})$, and Grover enhanced a brute force search in a database with $N$ entries such that it takes $\sqrt{N}$ operations.
Thus, Shor's algorithm weakens RSA, Diffie-Hellman, ECC, and any cryptosystem that relies on the aforementioned factoring and discrete-log problems, while Grover's algorithm greatly improves the attackers' efficiency in tasks such as password cracking, see Figure~\ref{fig:taxonomy-crypto}. 
Therefore, the design of cryptographic schemes that rely on the factoring and discrete logarithm problems should be revisited.  

There are, however, protocols that remain ``quantum-proof'' such as lattice-based cryptography, code-based cryptography, and multivariate cryptography~\citep{chen2016report,campagna2015quantum,perlner2009quantum}. 
In order to fully enjoy the advantages associated with these protocols, huge obstacles to making them practical solutions should be overcome. 
First and foremost, the adversary model in the realm of post-quantum cryptography should be well defined. 
The fact that adversaries might benefit from quantum technology, even in the future, makes it impossible to neglect the importance of defining precise adversary models~\citep{hsu2019how}. 
The second issue is related to determining the level of security that we expect from a cryptosystem. 
This level can be further translated to the key length, the time required to compromise the security of this key, and the time needed to implement a system to offer this level of security~\citep{mosca2018cybersecurity}. 
Last but not least, in line with the latter problem, post-quantum cryptographic schemes should be developed that achieve high efficiency and security simultaneously, see, e.g.,~\citep{nist2019}. 
The next section is devoted to this matter.

\begin{figure*}[t]
    \centering
    \includegraphics[width=\textwidth]{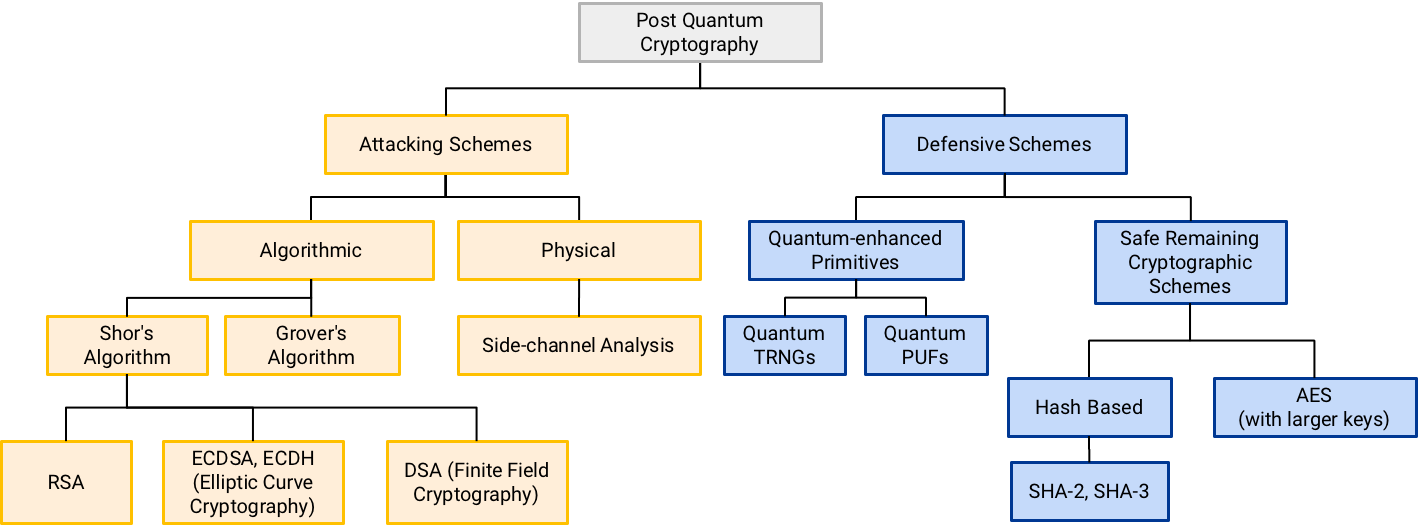}
    \caption{ 
    Taxonomy of post quantum-cryptographic schemes. 
    While quantum computation offers advantages in several useful applications, it can be misused by adversaries attempting to break the security of cryptographic primitives. In this regard, compared with the best classical computers, quantum computers facilitate running algorithms and conducting analysis, e.g., side-channel analysis. 
    }
    \label{fig:taxonomy-crypto}
\end{figure*}

\subsection{Benefits of Quantum Computing for Cryptography}
The danger of quantum computing is much publicly advertised; however, quantum technology is not simply restricted to methods of attacks. 
The quantum technology has two further useful properties in the context of cryptography: truly random processes and tamper-evident states.  
This positive side of this technology can be traced back to the introduction of two-state quantum systems, whose quantum states can be seen as the quantum superposition of two independent, physically distinguishable quantum states~\citep{leighton1965feynman}. 
This behavior of quantum states has been validated experimentally, e.g., in the Stern and Gerlach experiment performed in 1922, which explained the quantum property of spin in an electron~\citep{gerlach1922experimentelle}. 
The existence of such two-state systems has an important implication as they can also be seen as qubits, the basic unit of data generated using quantum technology. 
It must be noted that though a qubit may exist in a combination of states, yet after measurement, it results in any one of the two possible outcomes. This result remains unchanged even after repeated measurements as long as the procedure of measurement remains unchanged. 

This should not be confused with the concept of ``entanglement'' referring to a particle with individual states that cannot be defined independently, i.e., depending on the state of other particles- even if being spread far apart~\citep{leighton1965feynman}.  
This means that measurements on the state of one entangled particle affect the state of all of its entangled particles. 
Interestingly enough, this characteristic has found applications in quantum cryptography, particularly quantum key distribution. 
More concretely, in 1984, Charles H. Bennett and Gilles Brassard theorized a way of information-theoretically secure communication in a quantum system~\citep{bennett1984proceedings}. 
Their protocol, called BB84, has proposed a quantum key distribution (QKD), where one can create a key from qubits and transmit it to another user. 

Such a quantum key has some important properties: first, it is harvested from a truly random source because of the randomness inherent in the measurement of unpolarized qubits along an axis.  Secondly, if an eavesdropper attempts to read a quantum key, the interaction with the key's state is guaranteed to cause noticeable change, therefore providing verification that the key is secure -- something which is proven mathematically (a corollary of the no-cloning theorem~\citep{wootters1982single}), and is not possible in classical computing. 

Let us consider this protocol as an example of how quantum-enhanced cryptosystems can be realized in practice, thereby making the following observations.  
First, implementation of such a QKD system suffers from a stability problem -- qubits by nature are very unstable. Hence, if the system is not robust enough, it cannot be said with certainty whether a change in state was due to eavesdropping or system instability.
This issue has been studied and resolved (to some extent), see, e.g.,~\citep{asaad2020coherent}, resulting in further advancement in quantum cryptosystems~\citep{wallden2019cyber,dowling2003quantum}. 
Second, similar to various other types of cryptosystems, a QKD relies on the quality of the randomness source, from which the keys are extracted. 
Validation of this assumption is absolutely vital; otherwise, keys used in such a system can be prone to attacks (e.g., guessing). 
As a prime example of this, the security and applicability of a cryptosystem presented in~\citep{di2019perfect} has been questioned, partly due to the randomness source suggested in that study~\citep{hsu2020new}. 
Therefore, an important goal for future research is to sharpen our understanding of the conditions that ensure adequate security so that quantum cryptosystems can achieve a high level of reliability and security. 
In this respect, it is critical that the objectives of physical security are also achieved for such cryptosystems, e.g., quantum bit commitment, quantum coin-flipping, quantum fingerprints, quantum data hiding, quantum authentication, and encryption. 
These applications share some commonalities: (1) for them, the quantum technology has inherent advantages over classical protocols, and (2) they may need secure key generation and storage as well as secure execution. 
The next sections (Section~\ref{sec:pufs}-\ref{sec:sca}) describe methods and apparatus developed for this purpose. 
}
\section{Side-Channel Analysis}\label{sec:sca}
\sloppypar{
\begin{figure*}[t]
    \centering
     \includegraphics[width=\textwidth]{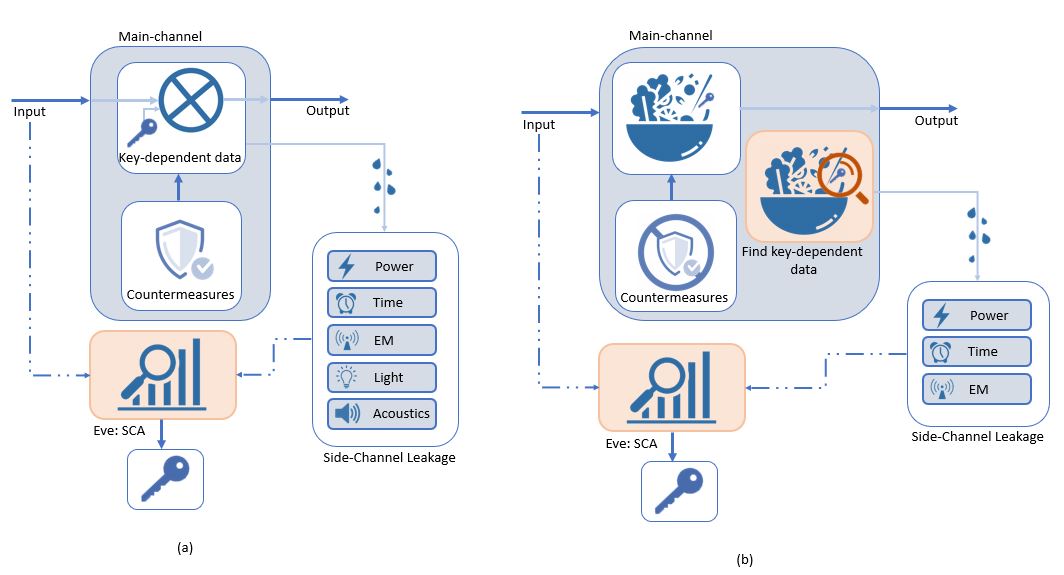}
     \caption{Comparison between side-channel attacks on (a) classical cryptographic algorithms, (b) post-quantum cryptographic algorithms. Adversary (Eve) needs to be able to control the circuit and has access to the inputs and data leakage. In (a) classical cryptographic algorithms dependency between the secret key and input data is more apparent, compared to (b) post-quantum cryptographic algorithms, in which adversary must put additional effort into finding the key-dependent data needed for SCA analysis.}
     \label{fig:sca-att}
\end{figure*}

Side-channel attacks (SCA) have been a prominent method of extracting sensitive data from cryptographic elements of the chip in the classic world. Such attacks exploit physical vulnerabilities in the hardware implementation rather than flaws of the mathematical structure of the algorithm. Compared to cryptographic attacks, which are modeled as black-box attacks, adversaries during SCA have access to the ``grey'' box~\citep{ref1}, in which internal physical quantities are observed and analyzed for the key extraction~\citep{ref1}. Side-channel is performed in two steps. Firstly, physical leakage of each query performed on cryptographic implementation needs to be turned into probability and score vectors~\citep{ref5}. This information is valuable because further key extraction can be performed. The second step is to sort information and search over every individual key until the entire key is completed and extracted~\citep{ref5,taha2015implementation}. The more complex or noisy the leaky data is, the more difficult the side-channel attack becomes to perform~\citep{ref5}.

\begin{figure*}[t]
\centering
  \includegraphics[width=\linewidth]{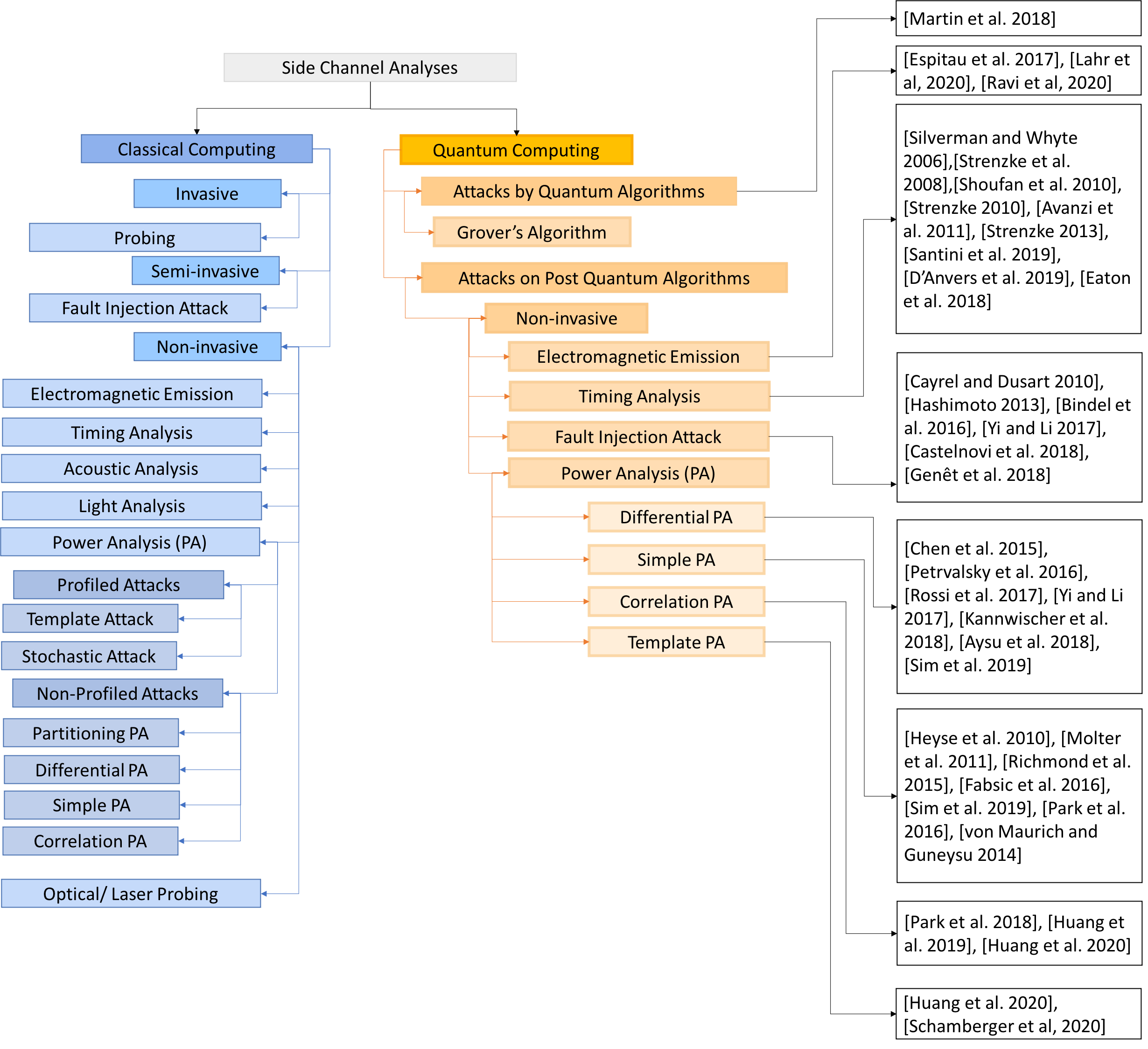}
  \caption{Taxonomy of side-channel analyses in classical and quantum computing settings. Physical side channel attacks performed on classical algorithms are shown in blue, and attacks on quantum-resistant algorithms are shown in orange, as well as the attacks by the quantum algorithms}
  \label{fig:sca-taxy}
\end{figure*}

\subsection{SCA in Classic World}

SCA and its countermeasures belong to a mature field, which has been investigated for more than twenty years~\citep{ref1}. A high-level overview of SCA has been reported in~\citep{ref3}. Further, numerous literature overviews of side-channel attacks have been reported over the years, such as~\citep{ref1,ref2,ref3,ref6,ref7}. SCAs are mainly separated into two categories based on the type of exploited information: physical and logical. Physical attacks obtain information found from physical features of the device~\citep{ref3}, such as power consumption and electromagnetic emissions. In contrast, logical attacks gain information from running software properties, such as data-usage statistics~\citep{ref3}, and data footprint, which can be exploited through cold boot attacks~\citep{ref14} (see, Figure~\ref{fig:rot}). 

This survey will focus on physical side-channel attacks, which can be categorized into invasive, semi-invasive, and non-invasive attacks, as shown in Figure~\ref{fig:sca-taxy}. The invasive side-channel attack destroys the physical packaging of the integrated circuit while maintaining functionality, and it cannot be returned to its original state. Semi-invasive attacks require backside decapsulation, in which attacker can perform photonic analysis~\citep{tajik2017power}, optical contactless probing~\citep{tajik2017photonic} or laser stimulation~\citep{lohrke2018key}. Finally, by performing non-invasive attacks, the attacker only observes specific physical parameters produced while the system is running, without affecting the IC packing. As discussed in the last section, such attacks can be active or passive.


The most abundant attacks reported are non-invasive attacks. While there have been ``exotic'' means of executing this attack, by using acoustics~\citep{ref13},~\citep{ref7},~\citep{ref8} or light~\citep{ref9} to produce useful data, the most versatile parameters to exploit the security of the chip are timing, power consumption, and electromagnetic (EM) emissions~\citep{ref7}. Power Analysis (PA) is a pioneering method in SCA, and most of the techniques used to analyze power emission data can be applied to data collected through EM. The taxonomy in Figure~\ref{fig:sca-taxy} also shows the categorization of these attacks into profiled and non-profiled attacks~\citep{ref2}. Profiling attacks, which model the device implementation, through either analysis of a large number of signals from a reference device in an SCA attack called Template attack, or through the pre-defined noise model with pre-defined function, as in Stochastic attack~\citep{ref2}, or Linear Regression attack~\citep{ref10}.  Non-profiling attacks such as Partitioning PA, Differential PA, Simple PA, and Correlation PA~\citep{ref2}, do not rely on modeling a reference device. 
The main building blocks of the attacks mentioned above are illustrated in Figure~\ref{fig:sca-att}. 
As depicted in this figure, SCAs performed in classic and quantum worlds share various similarities; however, due to their differences in nature, SCA in the quantum world should be considered in further detail, as explained below. 

\subsection{SCA in Quantum World}
While the development and design of SCA and countermeasures against SCA on pre-quantum classical algorithms have been the most explored in this field, SCAs in the quantum world have been researched on two fronts. Firstly, side-channel attacks in the quantum world became a point of interest for many researchers after the 2016 National Institute of Standards and Technology (NIST) call for post-quantum algorithms resistant to quantum computer attacks. Their susceptibility to side-channel attacks has been investigated due to the development of quantum algorithms. Another front is using the runtime and space usage advantages of quantum algorithm speed-up from quantum computers,~\citep{ref15}, to launch side-channel attacks on classical computation. The following sections will provide a detailed overview of post-quantum algorithms, which will be followed by side-channel attacks on them.

\subsubsection{Post-quantum Cryptographic Algorithms} 
\sloppypar{
Today's core cryptosystems, including public-key encryption, digital signatures, and key exchange, are mainly based on Diffie-Hellman key exchange, RSA (Rivest--Shamir--Adleman) public-key cryptosystem, DSA (Digital Signature Algorithm) and elliptic curve cryptosystems~\citep{ref17}. Since the development of quantum algorithms provide at least quadratic speed-up of computation, and at most exponential speed-up, RSA, DSA, and elliptic curve cryptosystems will no longer be secure after the creation of large size quantum computer~\citep{ref17}. However, symmetric encryption and hash functions will remain safe in the post-quantum era because exponential speed-ups for quantum search algorithms are not achievable against them~\citep{ref17}. The algorithms most promising against quantum computing are based on the lattice, code, hash, and multivariate public key problems. These algorithms will be introduced below. 
}

\vspace{0.5ex}

\noindent\textbf{Lattice-based algorithms} are considered promising due to their security under worst-case hardness assumptions and their simple construction. They are based on the shortest vector problem (SVP) in the lattice, as well as ring learning with error (R-LWE) problem, which is believed to be reducible to SVP, an NP-hard problem. 
Examples of Lattice-based algorithms are NTRU, BLISS, ring-TESLA, the GLP, and LAC~\citep{refZ,ref22,ref21}, which are based on ring learning with errors (R-LWE), the ring short integer solution (R-SIS), and the decisional compact knapsack (DCK) problem~\citep{refZ}. 
The original NTRU algorithm, consisting of algorithms NTRUSign and NTRUEncrypt, turned out to be susceptible to various attacks~\citep{ref22}, but its variation NTRU Prime is still a candidate in the second round of NIST standardization. Another lattice-based algorithm based on the Mersenne Low Hamming Combination Assumption, which involves error-correcting code, is Ramstake scheme~\citep{ref21}.

NTRU operations are based on truncated polynomial rings. Compared to the original NTRU, which consists of three stages (key generation, encryption, and decryption), there are two mechanisms for encryption in NTRU Prime: Streamlined NTRU Prime and NTRU LPRime, where latter one shares similarities with R-LWE schemes. NTRU Prime is exploited through the leakage in polynomial multiplication of private key and known ciphertext in the product scanning method~\citep{refX}. The original version of the NTRU algorithm was proven to be insecure because of the vulnerabilities exploited from the use of hash functions in encryption/decryption phases. 

Learning with errors based algorithms, e.g., FRODO, and R-LWE NewHope algorithms, have been attacked by differential power analysis. These attacks rely on a mathematically hard problem where the goal is to distinguish between a uniformly random sample from learning with error samples. Algorithm BLISS, which partially depends on R-LWE, has been exploited through electromagnetic emission analysis~\citep{refZ} and fault injection attack~\citep{ref22}. In addition to BLISS, ring-TESLA and the GLP have also been attacked by fault injection~\citep{ref22}. LAC algorithm is attacked by a timing attack in~\citep{ref21}. In~\citep{ref20}, in algorithms, FRODO and NewHope intermediate values of matrix and polynomial multiplications depend on the sub-keys. Intermediate values in polynomial multiplication in NewHope depend on the same coefficients of the secret polynomial, while in FRODO, intermediate values in matrix multiplication depend on the values from the secret matrix. BLISS, ring-TESLA, and GLP contain an important step of rejection sampling, which creates a distribution of created signatures independent of the secret key. All three algorithms share the following flow: firstly, the secret key and other variables are sampled and manipulated into the public key. The plaintext is hashed with the public key, either directly, or requires additional manipulation. The signature created consists of two polynomials. Rejection sampling is applied by compressing one polynomial, creating a signature independent of the secret key. In the decryption phase, the size of compressed and uncompressed polynomials is checked if they are equal, as well as the equality of the un-hashed value of polynomials and ciphertext.  

LAC algorithm consists of a key encapsulation mechanism (KEM) and public-key encryption (PKE). PKE consists of key generation algorithm, decryption, and encryption. In KEM, the key is created by expanding seed into a polynomial, which is uniformly and randomly sampled from a pseudo-random generator. Error correction capabilities of this scheme depend on the deterministic derivation of error-correcting values from the uniform and random seed. Ramstake algorithm depends on Mersenne prime numbers, which do not increase Hamming Weight when modulo operations are performed on them.

\vspace{0.5ex}

\noindent\textbf{Code-based McEliece scheme}, which has not been mathematically broken since its introduction in the 1970s, is based on the hardness of decoding a random linear error-correcting code, such as Syndrome Decoding, known to be NP-hard~\citep{singh2019code}. It requires a large key size, and it has been primarily used in encryption. Multiple optimized McEliece schemes have been demonstrated with the primary goal of achieving faster encryption~\citep{1468309}, such as quasi-cyclic low- and moderate- density parity code (QC-LDPC and QC-MDPC, respectively) in the McEliece scheme. Its difficulty is based on the Syndrome Decoding problem and the Goppa Codes Distinguishing problem. 

In classical McEliece, the private key is created from a parity check matrix, ``scrambling'' matrix, and permutation matrix. Encryption is done by adding a vector of errors to the manipulated plaintext. The first step of decryption is the creation of codeword, which is done by multiplying the ciphertext and permutation matrix. Then, further decryption is done by the Patterson algorithm, which computes the syndrome of the codeword. The syndrome is created by multiplying codeword with the transpose of the parity check matrix. Multiplication is done in two steps. Firstly, the syndrome is initialized as a vector of zeros. Then, the algorithm iterates through codewords. For i\textsuperscript{th} entry of codeword with of value of 1, i\textsuperscript{th} row of transposed parity check matrix is added to syndrome vector. The next step is to transform the syndrome vector into the syndrome polynomial. Patterson algorithm for that operation uses an algorithm that finds roots in polynomial and Extended Euclidean Algorithm (XGCD). Finally, the plaintext is obtained by multiplying syndrome with "scrambling" matrix, and by solving the key equation.

In QC-LDPC and QC-MDPC McEliece, the private key does not contain a permutation matrix, but it contains matrix Q created of a small number of 1s in every row. A code like this does not have an algebraic structure. With the sparse parity check matrix, error correction in decryption is efficient~\citep{ref409}. During the decryption phase in QC McEliece, the bit-flipping algorithm is used to create syndrome by computing the number of unsatisfied parity-check equations associated with every bit of parity-check matrix. Each bit that is involved in the number of equations greater than the threshold is flipped recomputing the syndrome, which gets recomputed until syndrome becomes zero. However, if in implemented design, the algorithm stops after a certain number of iteration, decoding failure happens.

\vspace{0.5ex}

\noindent\textbf{Hash-based algorithms} are used for digital signatures, and their security relies on the security of their associated hash function and/or binary hash tree structure~\citep{ref43}. A binary hash tree structure combines multiple one-time signature key pairs, and it can be stateful or stateless, depending on if the secret key gets updated or not after the signing. XMSS is a stateful digital signature scheme that is being standardized~\citep{ref43}. During the key generation of XMSS and tree generation, the public key is created from the seed coming from a PRNG. The signature is performed using W-OTS\textsuperscript{+} signature scheme. Compared to XMSS, SPHINCS is a stateless scheme, which in addition to W-OTS\textsuperscript{+}, also uses HORST signature schemes. Signatures are pseudo-randomly selected to sign the message~\citep{ref43}.
Side-channel attacks on hash-based post-quantum algorithms are rarely performed, compared to lattice- and code-based schemes, but work in~\citep{ref43} and~\citep{ref40} proposed DPA and fault injection side-channel attacks, respectively. 

\vspace{0.5ex}

\noindent\textbf{Multivariate Public Key Crypto-algorithms (MPKC)} rely on the NP-hard mathematical problem of solving a set of multivariate quadratic polynomial equations in a finite field. Various MPKC schemes have been proposed, but the most promising ones which are shown to be the fastest come from the step-wise triangular system family~\citep{ref42}: Rainbow, Unbalanced Oil and Vinegar (UOV), Tame Transformation Signature (TTS) and its enhanced version enTTS. 
Digital signature scheme enTTS is believed to be the fastest, which works with 20-byte hashes and 28-byte signatures in GF(2\textsuperscript{8}), as reported in~\citep{ref42}. The main building blocks of enTTS are secret multivariate polynomials of small size and linear maps. Most coefficients are zero, in which monomials do not occur twice. The central linear map consists of three layers~\citep{refTTS}. The hashed message needs to go through the computation of affine transformations (matrix-vector multiplications and vector additions), evaluation of polynomials (element multiplications), and solving of a system of linear equations~\citep{ref42}, to generate a signature of enTTS.

\subsubsection{SCA on Post-quantum Cryptographic Schemes}
As shown in Table~\ref{tab:tableSCA}, side-channel attacks in the quantum world are non-invasive attacks, and they exploit power leakage, electromagnetic emissions, and timing leakages. Fault analysis (FA) attack launched in the quantum world has been shown as both passive and active attack, in which an attacker actively changes and observes the behavior of the system, or just passively observes. Power analysis (PA) side-channel attacks launched in the quantum world are differential PA, correlation PA, and simple PA. 

\begin{center}
\scriptsize
\begin{table*}
\centering
\caption{Summary of physical attacks on post-quantum cryptographic algorithms (PQCA), where TA is Timing Attack, FA is Fault Analysis, PA is Power Analysis, SPA is Simple PA, DPA is Differential PA, CPA is Correlation PA, OTA is Online Template Attack and EMA is Electromagnetic Emissions Analysis }
\label{tab:tableSCA}
\begin{tabular}{|p{2.4cm}|p{1.5cm}|p{3cm}|p{5.2cm}| }
\hline
PQCA  & Type of SCA & Reference & Degree of Success \\
\hline
\multicolumn{3}{l}{\textbf{Code-Based Algorithm}}\\\hline
\multirow{11}*{McEliece}  & \multirow{6}*{TA} &\citep{ref45a}& Theoretical analysis of attack on degree of error locator polynomial.\\\cline{3-4}
& &\citep{ref45}& Theoretical analysis of secret permutation which decrease cost in brute force secret key recovery and proof of concept implementation. \\\cline{3-4}

& &\citep{ref46}& Experimental ciphertext recovery. \\\cline{3-4}

& &\citep{ref27}& Improved theoretical analysis of ~\citep{ref45a}.\\\cline{3-4}

& &\citep{ref44}& Experimentally recovered secret information: zero-element, linear and cubic equations.\\\cline{2-4}

            & FA &\citep{ref33}& Theoretical analysis of fault injection sensitivity.\\\cline{2-4}
            
            & \multirow{4}*{SPA} &\citep{ref24}& First analysis of experimental Goppa polynomials recovery required for secret key extraction from 8-bit AVR microprocessor.  \\\cline{3-4}
            
& &\citep{ref25}& Experimental ciphertext recovery from FPGA XGCD algorithm implementation.\\\cline{3-4}

& &\citep{ref30}& Experimental 80-bit private key and secret message recovery from STM32F4 Discovery Board and Atmel AVR XMEGA-A1 Xplained Board implementations. \\\cline{3-4}
& &\citep{ref29}& Experimental analysis of matrix multiplication implemented on ARM Cortex-M3 which is required for permutation matrix recovery in syndrome computation.\\\cline{2-4}

            & DPA &\citep{ref31}& Experimental recovery of 64 by 64 permutation matrix from the ARM Cortex-M3.\\\hline
\multirow{7}*{QC-LDPC/MDPC} & \multirow{2}*{TA} & \citep{refY}&Theoretical model of partial key recovery. \\\cline{3-4}
& & \citep{ref37} & Theoretical and experimental analysis of key recovery which does not depend on decoding failure rate using 2\textsuperscript{20} samples for 80-bit key, 2\textsuperscript{23} samples for 128-bit key and \textsuperscript{25} for 256-bit key. \\\cline{2-4}
& \multirow{3}*{DPA} & \citep{ref32}& Experimental full key recovery from FPGA implementation presented at Design, Automation and Test in Europe Conference 2014.\\\cline{3-4}
& & \citep{ref38} & Experimental partial key recovery from ChipWhisperer evaluation platform followed by entire key recovery computed by solving the system of noisy binary linear equations. \\\cline{3-4}
& & \citep{ref36} & Experimental full key recovery from 32-bit processor eliminating need for solving linear equations from~\citep{ref38}. \\\cline{2-4}
& \multirow{2}*{SPA} & \citep{ref35}&Theoretical full key recovery analysis. \\\cline{3-4}
& & \citep{ref36} & Experimental full key recovery from 32-bit processor breaking the countermeasure proposed by~\citep{ref38}.  \\
\hline \multicolumn{4}{r}{\textit{Continued on next page}} \\
\end{tabular}
\end{table*}

\begin{table*}
\centering
\begin{tabular}{|p{1.8cm}|p{1.5cm}|p{3cm}|p{5.2cm}| }
\multicolumn{4}{c}%
{\tablename\ \thetable\ -- \textit{Continued from previous page}} \\
\hline
\multicolumn{3}{l}{\textbf{Lattice-Based Algorithm}}\\\hline
\multirow{3}*{NTRU Prime} &  \multirow{2}*{CPA}  &\citep{ref18} & Theoretical and practical analysis of polynomial multiplication needed for full key recovery implemented on STM32F303RCT7 32-bit microcontroller.\\\cline{3-4}
& & \citep{refX} & Experimental secret key recovery from the polynomial multiplication from Cortex-M4 implementations.\\\cline{2-4}
            & OTA & \citep{refX}& Experimental recovery of full private key from the Cortex-M4 board implementation.\\\hline
            
NTRUEncrypt & TA & \citep{ref19} & Theoretical and experimental analysis of partial secret key recovery.\\\hline

NewHope, FRODO & \multirow{2}*{SPA} &\citep{ref34}& Experimental full secret key recovery from R-LWE-based schemes implemented on 8-bit microcontroller.\\\cline{2-4}

&DPA &\citep{ref20}& Experimental full secret key recovery from SAKURA-G FPGA Board implementation with 99\% success rate.\\\hline

LAC & TA &\citep{ref21}& Experimental analysis of full secret key recovery under 2 minutes using less than 2\textsuperscript{16} queries.\\\hline
\multirow{2}*{BLISS} & EMA & \citep{ref22}&Experimental full secret key recovery on embedded 8-bit AVR implementation.\\\cline{2-4}
            & FA &  \citep{refZ}&Theoretical fault sensitivity analysis of implemented algorithm.\\\hline
ring-TESLA, GLP & FA &\citep{refZ}&Theoretical fault sensitivity analysis of implemented algorithms.\\\hline
Ramstake & TA & \citep{ref21}& Experimental full secret key recovery under 2 minutes using approximately 2400 decryption queries.\\\hline
\multicolumn{3}{l}{\textbf{Hash-Based Algorithm}}\\\hline
\multirow{3}*{SPHINCS} & DPA &\citep{ref43} & Theoretical partial secret key recovery from simulated implementation.\\\cline{2-4}
            & \multirow{2}*{FA} &\citep{ref40} & Theoretical partial secret key recovery from multiple compelled signatures \\\cline{3-4}
& &\citep{ref39}& Experimental analysis of theoretical attack from \citep{ref40} implemented on Atmel ARM-based SAM3X8E with over 30\% success probability from 64 forgery attempts.\\\hline
XMSS & DPA &  \citep{ref43} & Unsuccessful theoretical key recovery.\\\hline
\multicolumn{3}{l}{\textbf{MPKC-Based Algorithm}}\\\hline
\multirow{2}*{UOV, Rainbow} & FA &  \citep{ref49} &Theoretical analysis of partial key recovery.\\\cline{2-4}
            & CPA &\citep{ref41} &Experimental full secret key recovery from 8-bit AVR microcontroller with help of algebraic key recovery attack.\\\hline
TTS & FA &\citep{ref49} & Theoretical analysis of partial key recovery.\\\hline
enTTS & DPA, FA &\citep{ref42}& Theoretical analysis of partial key recovery from \textit{naive} Application Specific Integrated Circuits (ASIC) implementation. \\\hline

\end{tabular}
\end{table*}
\end{center}

%
\vspace{0.5ex}

\noindent\textbf{Differential Power Analysis (DPA)} is a statistical attack which analyzes measured power consumption from traces of cryptographic algorithm implementation. Attacked traces are intermediate values that are manipulated in a way that they can be expressed as a function of the secret key and known value. The attacker often uses Hamming Weight (HW) or Hamming Distance (HD) model to predict power consumption. HD models leakage as switching bits, while HW models leakage based on the number of bits in measured data. The attackers use a statistical tool to compute the correlation between predictions and the acquired power consumption traces. DPA vulnerabilities of the post-quantum crypto schemes are discussed below, and they are summarized in Table~\ref{tab:tableSCA}.

\begin{itemize}[leftmargin=0.25cm]
\item \textit{Code-based algorithms}: DPA usually attacks an intermediate value, which is a function of known data and secret key. This dependency is not straightforward in the classical McEliece scheme. The syndrome was thought to be the variable needed to be attacked, but ciphertext only directs how the syndrome is computed for the parity check matrix and does not contain the ciphertext in it.
DPA attack on code-based scheme classical McEliece from~\citep{ref31} is launched on bit permutation of ciphertext. HW model is applied to individual bits of leakage model. Correlation analysis must be performed for each input bit. The permutation matrix gets recovered by comparing known ciphertext and permuted ciphertext to correlation peaks from each measurement. One measurement corresponds to one row of the permutation matrix. However, in  QC McEliece schemes, DPA is performed on syndrome computation, because the parity matrix is sparse, containing the same information in each row rotated one bit each row,~\citep{ref32}. 

DPA methodology on constant-time multiplication~\citep{ref36}, which forms syndrome, consists of splitting the attack position into two parts, word rotation and bit rotation, which computes different parts of the parity matrix. Constant-time masked multiplication has been implemented as a countermeasure against timing attacks. However, this countermeasure is vulnerable against DPA in private syndrome computation. Prior to~\citep{ref36}, attacks failed to recover the entire key because constant-time multiplication software implementation results were saved in the same register, creating numerous candidates for secret indices. Further, the system of linear equations had to be solved, such as in~\citep{ref38}, whose complexity increases with the number of possible candidates. However, the attack in~\citep{ref36} proposed multiple and single trace attacks to overcome the need for solving linear equations, as stated in Table~\ref{tab:tableSCA}. If the process provides single-bit shift instructions, a single trace attack is sufficient by recognizing left shift instruction, while for multiple bit shift instructions, multiple track attack is performed to recover correct secret indices from a numerous number of possible candidates. 

Multiple-trace DPA methodology consists of splitting the attack position into two parts. In the first part, masking is performed with all ones or all zeroes. In this step, the power consumption at each point is modeled as a sum of data-dependent power and Gaussian noise power. Modeling the data-dependent power consumption with HW, there is a linear relationship between total power consumption and HW modeled value. Using the Pearson correlation coefficient between these two parameters, the first part of the correct indices can be recovered. In the second part, the CPA is performed to find the rest of the correct indices by bit rotation. 

\item \textit{MPKC algorithms}: DPA on MPKC recovers the secret affine map by targeting matrix-vector product, such as in Rainbow and UOV schemes in~\citep{ref41}. The hurdle present in this attack is that neither intermediate values nor the vector which multiplies secret affine maps are known. However, if the scheme is implemented with a specific key structure~\citep{ref41}, CPA can be performed to extract the values of the secret affine map. \citep{ref41} has launched the attack on 8-bit AVR microcontroller, recovering full secret key, see Table~\ref{tab:tableSCA}. DPA attack on enTTS algorithm requires help from fault analysis, to perform the DPA attack correctly.~\citep{ref42} analyzed a DPA attack on enTTS by executing algorithm several times implemented on Application-Specific Integrated Circuits (ASIC), with different inputs and getting a set of power consumption traces from affine transformations and central map with multivariate polynomials. While the algorithm is executing, sensitive variables are manipulated by fault attack, which changes the random values to fixed values. This sensitive variable is related to the secret key and known variable. Making a hypothesis about the secret key, the attacker can predict the sensitive value and corresponding leakage. Correlation analysis is then performed between the predicted value and measured power. To map hypothetical sensitive values to corresponding leakages, the Hamming Distance model is used because enTTS is implemented on CMOS.

\item \textit{Lattice-based algorithms}: CPA on lattice-based algorithm NTRU Prime in~\citep{ref18} concentrates on multiplication of ciphertext and private key in the decryption phase. It considers HW of intermediate value to be expected power consumption. The correlation analysis algorithm processes all HW polynomials and the corresponding measured power values with respect to the user-set correlation threshold. The candidate is determined from all optimal guesses, with the largest absolute value of the correlation coefficient. The attack recovers full key from polynomial multiplication from the NTRU Prime implemented on the 32-bit microcontroller, \citep{ref18}, see Table~\ref{tab:tableSCA}.
Intermediate states of matrix-polynomial multiplication also depend heavily on the same subkey, in other lattice-based algorithms, such as FRODO and NewHope. The main limitation of applying DPA on lattice-based algorithms is the frequency of false positives because similar outputs are created for similar sub-keys. To mitigate this issue, the attacker observes intermediate results and key bit by bit. According to to~\citep{ref20}, because of modular reductions in the process, only one bit will have a high correlation. This attack in~\citep{ref20} forms an ensemble of possible keys. 

\item \textit{Hash-based algorithms}: Power analysis attack on hash-based algorithms requires a function which depends on the secret key and known value, to be found. Then, the function is called twice, during the key and signature generations. A limited number of function calls is the reason for a few SCA on hash-based algorithms being reported. Measurement is filled with noise, and in both function calls, a value known to the attacker is the same. The ideal case occurs when the hash function and PRNG do not leak any information, which is not entirely accurate in practice ~\citep{ref43}. Another difference is that the one-time signature is called multiple times during the authentication path, giving attackers a greater possibility to attack the scheme. In practice, the attacker is not powerful enough to arbitrarily choose the leakage function and change it during every signature and key creation, so authentication path leakage gets reduced to the leakage of a single, one-time signature scheme OTS. DPA has been analyzed on the SPHINCS algorithm, as reported in Table~\ref{tab:tableSCA}, while the same attack is not successful against the XMSS algorithm.
\end{itemize}

\noindent\textit{Comparison to classic world algorithms: }
DPA against AES consists of attacking each round of AES and using the recovered secret as a known variable in the next round,~\citep{refAES}. Compared to any of the attacks on post-quantum cryptographic algorithms, this attack is straightforward and does not require a search for a function that depends both on sub-key and value known to the attacker. In the classic world, various multi-bit statistical tools are used in final key recovery, while in the post-quantum world, correlation analysis seems only to be used, thus reducing DPA to CPA. 

\vspace{0.5ex}

\noindent\textbf{Simple Power Analysis (SPA):}
Simple power analysis (SPA) visually analyzes power traces measured over time. Such attack exploits key and data dependencies during the computation. It can also be successful against recognizing certain operations within the algorithm since each operation has its own power signature~\citep{ref24}. SPA can only be used when the signal-to-noise (SNR) ratio is high enough; otherwise, DPA would be a better choice. SPA on post-quantum cryptographic algorithms is reported on code and lattice-based schemes.

\begin{itemize}[leftmargin=0.25cm]
\item \textit{Code-based algorithms}: SPA performed on classical McEliece has been able to extract permutation and parity-check matrices. If these operations are computed individually, then matrices are successfully recovered. However, if these two are combined in computation, they are recovered combined as well~\citep{ref24}. The HW of permuted ciphertext and original ciphertext is the same because only the location of the ciphertext gets permuted. Permutation matrix can be recovered if the attacker only considers ciphertext with HW equal to 1. The attacker has also been able to visually recognize when the summation is performed~\citep{ref24,ref29}. SPA attack also recovers an error vector, which is added to encoded plaintext. The attacker needs to have access to the Euclidean algorithm to recognize the error locator polynomial,~\citep{ref25}. The decryption of chosen ciphertext, and measuring power traces, the attacker is able to recover the error vector from power peaks which correspond to Euclidean algorithm iteration numbers, see Table~\ref{tab:tableSCA}. 

\item \textit{Lattice-based algorithms}: SPA on R-LWE based cryptographic scheme attack the decryption phase by visually inspecting if the modular addition is larger than the modulus~\citep{ref34}. In the first step of the SPA on R-LWE, the ciphertext is chosen. Then decryption is performed on the chosen ciphertext. The next step is to recognize if the modular addition is executed, and if it is, the secret key can be recovered. SPA has been performed on NewHope and FRODO, experimentally recovering entire secret key with a high success rate, as noted in Table~\ref{tab:tableSCA}.
\end{itemize}

\noindent\textit{Comparison to classic world algorithms: }
SPA in the classic world is dangerous, having even been successful against RSA (by recovering a private bit from the square-and-multiply algorithm) and against KeeLog. Compared to the classic world, the post-quantum world does not have easily distinguishable functions, especially considering that entire matrices need to be recovered, compared to keystream in the classic world. 

\vspace{0.5ex}

\noindent\textbf{Electromagnetic Emission Attacks (EMA): }
Techniques used in electromagnetic emission analysis are the same as the ones used in power analysis. The only difference in attack is how data is acquired, as the measurement setup is different. 
The main challenge is to recognize the possible data leakage. In the post-quantum cryptographic algorithm, the EMA attack has been launched on the lattice-based algorithm BLISS recovering the full secret key from embedded 8-bit AVR implementation; see Table~\ref{tab:tableSCA}. The source of leakage in this algorithm is in the rejection sampling algorithm during the signature generation~\citep{ref22}. The attacker is assumed to know the absolute norm of integer of interest. The rejection sampling step leaks the relative norm of the secret key. The rejection sampling is important to achieve correct output distribution, and its construction is very similar to the square-and-multiply algorithm, which allows for visual inspection of bits through Single EMA (SEMA)~\citep{ref22}.  

The research in this direction has resulted in a generic attack presented in~\citep{ravi2020generic}, which can be adaptable to various LWE-based PKE/KEM schemes. 
The proposed attack is a chosen-ciphertext one enabled through EM side-channel analysis. 
The leakage is identified within the constant-time decoding procedures of error-correcting codes (ECC) that are employed to examine the validity of decrypted codewords. 
Similarly, the Fujisaki-Okamoto transform used to detect invalid or maliciously formed ciphertexts has shown to be susceptible to the proposed attack. 
Building upon this attack, in~\citep{schamberger2020power}, Schamberger et al. has proposed the first profiling power side-channel attack mounted on the KEM scheme used in code-based Hamming Quasi Cyclic (HQC) cryptosystem~\citep{melchor2018hamming}. 
Although the constant-time implementation of an ECC has been considered in~\citep{melchor2018hamming}, the ECC decoders of the reference implementation exhibit a power consumption pattern depending on whether an error must be corrected.
This has been leveraged in their attack to disclose the entire secret key. 

Another example of EMA is given in~\citep{lahr2020side}, where the system under attack is the code-based Niederreiter. 
The attack is based on timing side-channel plaintext-recovery attack introduced in~\citep{ref46}; however, as the constant-time hardware implementation is selected to mount the attack, EM side-channel is substituted for the timing side-channel. 
The attack has been further optimized in terms of the  number of required side-channel queries. 

\vspace{0.5ex}

\noindent\textbf{Timing Attacks (TA): }
Timing attacks (TA) are a type of side-channel attack in which the attacker exploits the time required for the completion of a logical operation. The time required for operation execution can also differ within the operation itself, based on the inputs being processed. Timing attacks in the post-quantum world have been successfully launched on code-based and lattice-based algorithms. 

\begin{itemize}[leftmargin=0.25cm]
    \item \textit{Code-based algorithms}:
The attack in~\citep{ref45a} is the first timing attack on the McEliece algorithm without any countermeasures exploiting the time needed for decryption execution and learning the dependence of errors in decoding algorithm and error locator polynomial. The attack is further extended and analyzed in~\citep{ref45}, where the private key in the Patterson algorithm is attacked by exploiting the time required to solve the key equation. It is accomplished by attacking private key in the Patterson algorithm, where error locator polynomial is related to the secret, and the time needed to solve the key equation leaks information about used polynomials. However, this attack was shown to be impractical in~\citep{ref44}. The timing attack in~\citep{ref44} takes advantage of the multiplication of syndrome with a scrambling matrix, more specifically of syndrome inversion. \citep{ref44} expands on work in~\citep{ref45} by using the leakage from syndrome inversion, together with the leakage in solving the key equation, creating a practical attack that recovers secret information: zero-element, linear and cubic equations, see Table~\ref{tab:tableSCA}.

Timing attacks against optimized versions of McEliece algorithm, QC-MDPC, and QC-LDPC, are proposed in~\citep{refY,ref37}. In~\citep{ref37}, the parts of the key are recovered by exploiting the decryption failure rate and measuring the number of iterations in the decryption phase. Average decryption time to correct errors varies, and~\citep{ref37} shows the correlation between errors and key. 

\item \textit{Lattice-based algorithms}:
A timing attack has been reported on lattice-based algorithms, specifically R-LWE schemes. In this attack,~\citep{ref21}, decryption errors are detected and exploited prior to being corrected. Because constant-time error-correcting schemes are not easily implemented, decryption takes different time for codewords, which those that contain and those that do not contain errors.
\end{itemize}

\noindent\textit{Comparison to classic World Algorithms: }
Timing attacks in quantum settings are significantly underdeveloped compared to numerous timing attacks in the classic world. Compromising the error correction also exists in the classic world, which is only one of the numerous techniques reported. Another difference is that countermeasures in the classic world are developed to the extent of questioning the practicality of timing attacks, while in post-quantum world timing attacks are still very strong, considering error-correction time varies~\citep{ref21}.

\vspace{0.5ex}

\noindent\textbf{Fault Attacks (FA):}
Errors in computation can be intentional or unintentional. Unintentional errors can occur due to poor quality of the component, fuzzy noise, or external natural effects, such as space radiation for systems launched into space. Chip designers take these possible errors into account when designing the chip. However, faults, which are intentionally introduced, can be a powerful attack pattern. The attacker makes the system behave incorrectly while observing the behavior, comparing it with correct behavior, and reverse engineering sensitive secret data. Fault injection attacks have been investigated in post-quantum cryptographic world on code-based algorithms in~\citep{ref33}, lattice-based algorithm in~\citep{refZ}, MPKC algorithm in~\citep{ref49}, and hash-based algorithm in~\citep{ref40}, as reported in Table~\ref{tab:tableSCA}.

\begin{itemize}[leftmargin=0.25cm]
\item \textit{Code-based algorithms}: According to~\citep{ref33} McEliece scheme is intrinsically resistant against fault injection attacks because of underlying error correction code. Theoretical fault analysis in~\citep{ref33} focuses on achieving denial of service by the introduction of random fault on one bit in two different parameters encoding the plaintext: addition of errors and multiplication with the public key. The third option evaluated is corrupting a plaintext. They show that only if the plaintext is corrupted, denial of service will be triggered, while for other cases underlying error correction code will correct the corruption. However, optimized QC-LDPC and QC-MDPC McEliece schemes are more sensitive to fault injection attacks because, in the sparse matrices, the faulty random bit gets re-used multiple times, greatly diffusing the fault.

\item \textit{MPKC algorithms}: Fault attacks on MPKC algorithms change the unknown coefficients of polynomials in the central quadratic map. For step-wise triangular systems, such as enTTS, the attacker can recover part of secret affine transformation directly by owning a message and signature, which is computed with a central map containing faults. In practice, this attack does not always succeed, because central quadratic map contains three parameters, two secret affine transformations, and central map~\citep{ref49}.  

Another interesting example of such attacks has been reported in~\citep{mus2020quantumhammer}, where Lifted UOV (LUOV) signature scheme has come under attack. 
For this, first faults have been injected through a software-only approach, namely Rowhammer attack~\citep{kim2014flipping}. 
Afterward, the faulty signatures are collected, and the divide and conquer attack leveraging the structure in the key generation part of LUOV. 
The proposed attack has successfully recovered all eleven thousand key bits in less than 4 hours of an active Rowhammer attack. 

\item \textit{Hash-based algorithms}: The first fault injection attack on hash-based algorithms was proposed in~\citep{ref40} against SPHINCS. The attack is performed in two steps: faulting and grafting. To execute this attack, the same message is signed twice, but the fault provoked in the second signing gives a different signature than the first. The grafting step is an analysis of two messages, in which parts of the secret key from one-time-signature (OTS) are recovered. The attacker grafts a tree of signatures from correct SPHINCS to the tree of signatures created from faulty OTS. This attack has been experimentally launched in~\citep{ref40} successfully recovering secret key with a probability of 30\%, as noted in Table~\ref{tab:tableSCA}.
\end{itemize}

\noindent\textit{Comparison to classic world algorithms:} Fault injection attacks in the classic world have received considerable investigation and can be classified into non-invasive and semi-invasive attacks, which require sample preparation and physically injecting faults into the system, through bombarding it with radiation or light. In post-quantum settings, all proposed attacks are still on the software level, as many algorithms are not implemented on a physical system. 
It can also be interesting to observe that these attacks have been primarily launched on classical cryptographic schemes, as shown in Fig.~\ref{fig:sca-diff}. 

\begin{figure*}
  \includegraphics[width=\linewidth,scale=0.5]{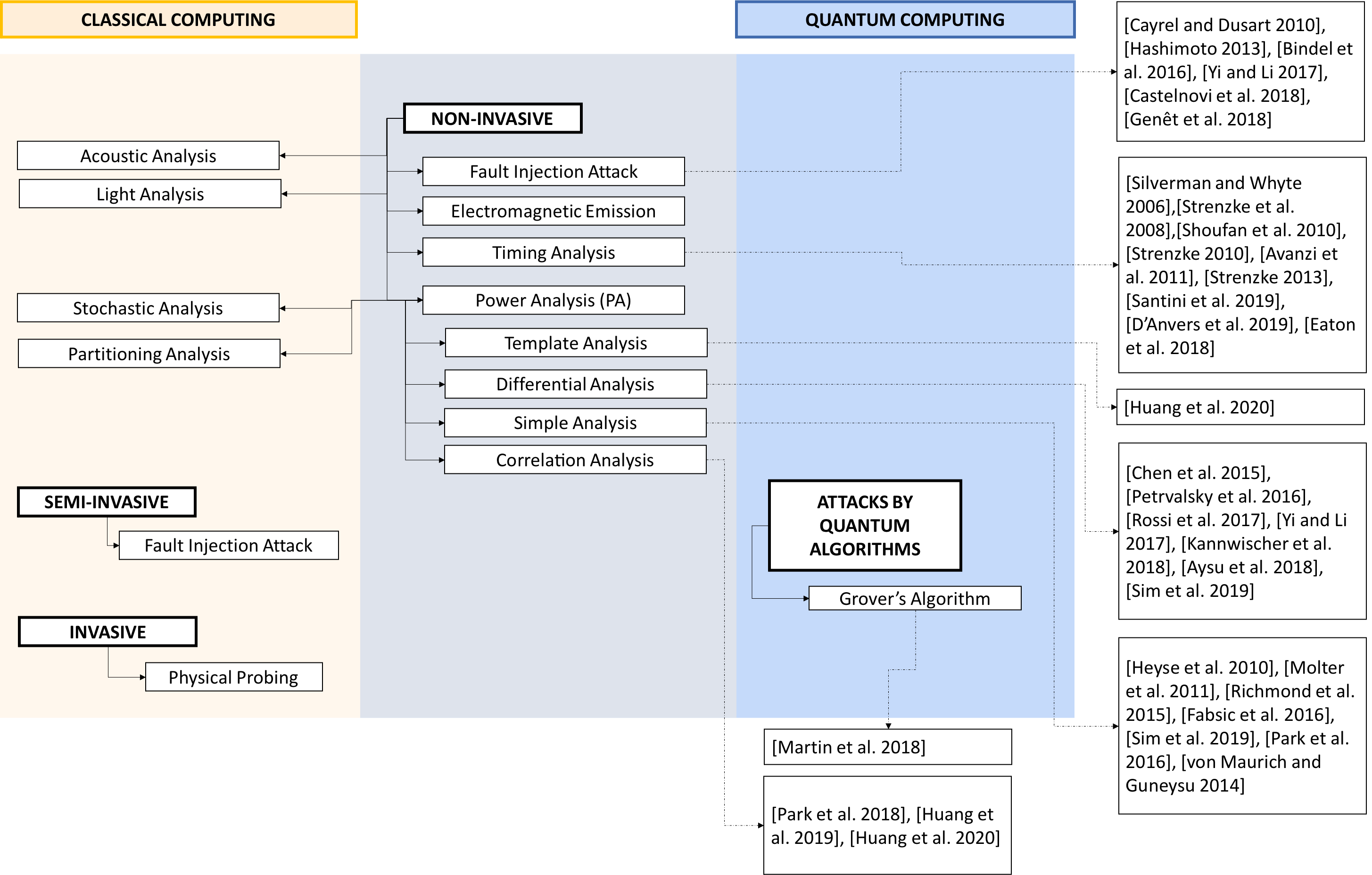}
  \caption{Differences and similarities in taxonomy of side channel attacks in quantum and classical computing.}
  \label{fig:sca-diff}
\end{figure*}

\subsubsection{SCA by Quantum Computer on Classical Algorithms}
After the development of Shor's algorithm over twenty years ago, which provides exponential computational speed-up, theoretical quantum algorithms have been greatly investigated~\citep{ref17}. Proposed algorithms are based on problems considered hard by computational complexity theory~\citep{ref15}. They have been reviewed by~\citep{ref15},~\citep{ref16} and reported by NIST~\citep{ref17}. It is hard to imagine the range of all applications of quantum algorithms and what spheres of lives would be improved with exponentially sped-up computations. However, in terms of side-channel attacks and privacy issues, the computer science problem of unstructured search~\citep{ref15} has the most highlights. According to to~\citep{ref15}, the unstructured search is a problem which evaluates function $f(x)$ and searches for $x$, if a solution exists, without any prior knowledge of the function. This is classified as NP-hard complexity problem~\citep{ref15} and in the worst-case scenario solution is found after $2^{n}$ function evaluations. The Quantum algorithm, Grover's algorithm, solves the problem in $\sqrt{2^{n}}$ number of iterations, providing quadratic speed-up compared to existing classical algorithms. Several works use quantum algorithms to perform brute force attack and advance classical crypto-analytics~\citep{ref50}, but there is only one work, the best to our knowledge, which uses Grover's quantum algorithm to investigate side-channel attacks performed by quantum computer~\citep{ref50}, even though side-channel leakage is considered as a quite structured set of data.

As discussed earlier, the side-channel attack consists of two steps. Firstly, data is obtained, which is processed into optimal sub-key candidates. Then, the search over those values is performed to find the most likely candidates for the construction of the entire private key. According to to~\citep{ref50}, the first step of side-channel attack remains the same, once quantum algorithms are implemented on quantum computers, because there are no benefits from quantum speed-up on the processing of already low complex data unless the attack is performed with few queries and collects large amounts of data, which is unlikely, even in hash-based algorithms. The second step, however, which analyzes data and search for possible keys, can benefit from quantum speed-up. The issue is that side-channel leakage data is highly structured~\citep{ref50} data, while Grover's algorithm operates on unstructured data~\citep{ref15}. According to~\citep{ref50}, this issue had already been investigated, in which Grover's algorithm, in addition to ordered data, also takes an advice distribution for the set. An additional issue is that data obtained through a side-channel attack is independent, and it is not sorted and ordered as Grover's needs. 

Work in~\citep{ref50} presents a novel version of the previous works in which the need to sort data in order of likelihood of finding a solution is considered, and that distribution advice is provided. This analysis is theoretical. Assuming that attacker has quantum RAM, the attacker can achieve quantum speed-up with a similar assessment of the time complexity. Attack~\citep{ref50} is recognized to be limited because it uses the principles of the classic world and applies them to quantum settings. However, this work puts the roots in a new research direction. Regarding quantum search, rank estimation has also been researched, and~\citep{ref52} shows provable poly-logarithmic time- and space- complexity. However, the same group that originally used Grover's algorithm to improve SCA~\citep{ref50} has been working on improving original concept, and in~\citep{ref51}, they draw the parallel between rank estimation and key enumeration techniques, which open doors for further research which should focus on optimizing the algorithm, so it does not have to sort an entire set of candidates prior to executing Grover's algorithm~\citep{ref52}.

}

\section{Random Number Generators}\label{sec:trngs}
\sloppypar{
One of the most important requirements for virtually all keyed cryptographic primitives is the existence of a random, unique, unpredictable key. 
TRNGs\footnote{Note that the scope of this paper does not cover the concept of algorithmic, cryptographically secure pseudorandom number generators, e.g.,~\citep{blum1984generate}. } play a crucial role in implementing practical schemes. 
Among such schemes, ones designed to ensure the physical security include, e.g., anti-side-channel masking schemes~\citep{shamir1979share,prouff2013masking}, secure entity authentication protocols~\citep{van2012reverse}, and intellectual property (IP) protection~\citep{9082137,roy2008epic,guin2016fortis}, just to name a few. 
TRNGs have considered promising candidates for generating keys due to their specific characteristics, including non-reprehensibility, and uniqueness (i.e., being instance-specific).  
However, these somewhat ``black-box'' assumptions about the security of TRNGs have been shown not to be valid in practice. 
More precisely, their information security is reduced to physical security, which is indeed broken by attacks mounted at a physical level, cf.~\citep{maes2013physically}. 

On the other hand, quantum counterparts of TRNGs, so-called quantum random number generators (QRNGs), have been advancing and becoming a well-established quantum technology with some provable assurances. 
In general, QRNGs can be seen as a special type of TRNG, in which quantum mechanical effects contribute to producing random numbers. 
Applications of QRNGs range from simulation to cryptography, and interestingly enough, commercial QRNGs have been offered since 2014, to the best of our knowledge cf.~\citep{herrero2017quantum,abellan2018future}. 
Regardless of their applications, similar to TRNGs, QRNGs must attain reliability\footnote{This should not be confused with the reliability of PUFs discussed in Section~\ref{sec:pufs}. Here, the reliability of TRNGs and QRNGs reflects the fact that they should be robust against tolerances in their components/environments.} as their failure can expand the attack surface embodying QRNGs. 
For instance, the BB84 quantum key distribution (QKD) protocol is vulnerable to attacks reported in the literature~\citep{bouda2012weak,li2015randomness}. 
This section covers such attacks and discusses how they could be a threat even in the post-quantum era. 

\begin{figure*}[t]
    \centering
    \includegraphics[width=0.8\textwidth]{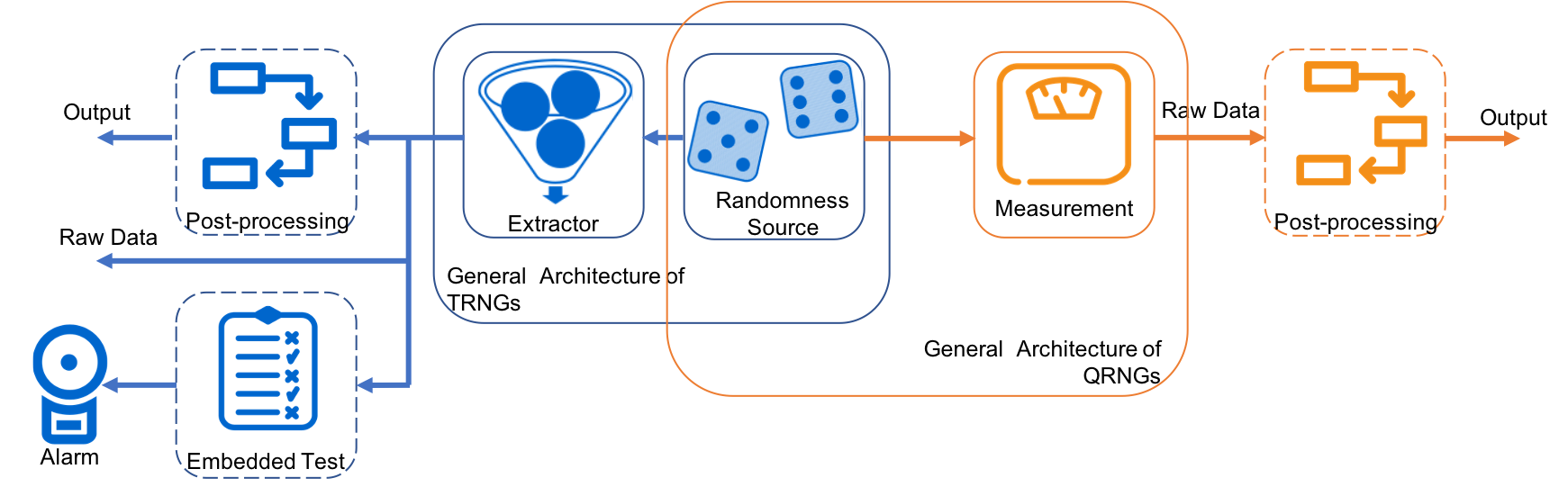}
    \caption{(Inspired by~\citep{bayon2016fault,herrero2017quantum}) Typical modules included in the design of a TRNG and QRNG. 
    As shown in this schematic, for TRNGs and QRNGs, the main building blocks comprising the general architecture can be considered similar. 
    More specifically, even if not mentioned explicitly for TRNGs, measurement of the random, physical variable should be carried out before applying the randomness extractors. 
    Post-processing is usually applied to improve the randomness while the embedded test is added by the designer in some cases. 
    In addition to running a randomness extraction algorithm, this step can be taken to buffer and accumulate samples before outputting the random strings and/or to test whether the generator is working properly cf.~\citep{schindler2002evaluation}. 
    }
    \label{fig:trng-blocks}
\end{figure*}
\subsection{TRNGs: Designs and Physical Attacks}\label{sec:trngs-design-attacks}
Generally speaking, it is expected that the output bits of a TRNG are uncorrelated and unbiased. 
This is indeed possible only in an ideal scenario, where the bias is handled carefully through, e.g., a randomness extractor, and TRNG is not susceptible to external influences. 
As these requirements cannot be (easily) fulfilled in practice, the output of a TRNG can often be predicted through bit-by-bit deduction attacks. 
To overcome this failure, the designers attempt to come up with architectures, whose outputs can be predicted only with a negligible probability. 
Nevertheless, it does not hold if the output of the TRNGs is, even to some extent, under the control of an adversary. 
For this purpose, the adversary can induce variations in the setting of the TRNG, including, e.g., manipulating input and/or output bits (e.g., introducing spikes in the power supply), emitting electromagnetic signals, changing the temperature outside the spec's operational range, etc. 
As described below, the common design of TRNGs may not resist against such attacks. 

Typically, the general architecture of a TRNG is composed of two main modules, namely a source of randomness and a randomness extractor. 
Here our main focus is on \emph{physical} TRNGs, built on the basis of physical phenomena, with the most prominent examples including noise-based~\citep{petrie2000noise} and free-running oscillators~\citep{kohlbrenner2004embedded}. 
This general architecture is usually implemented in a system along with a post-processing block (algorithmic and/ or cryptographic) and, possibly, an embedded testing one, see, Figure~\ref{fig:trng-blocks}. 
Obviously, these modules and blocks can be vulnerable to various types of attacks falling into two categories: invasive and non-invasive ones. 
In this context, being invasive means that the attack is powerful and usually can have a permanent effect on a circuit, e.g., through inserting a pin, burning a hole, etc.~\citep{sunar2006provably}. 
In our survey, if not mentioned otherwise, we review non-invasive attacks, where a TRNG is under attack temporarily, primarily, to introduce bias into its outputs. 

From another point of view, an attack can be categorized into passive and active attacks. 
In the former, the side channels such as the power consumption, execution time, or electromagnetic emanations are considered, which can be helpful to obtain some information regarding the internal functionality of the TRNG, e.g., frequencies, location, sampling frequency, etc. 
Passive attacks can target the randomness extractor, and the output of the system, before or after the post-processing model~\citep{bayon2016fault}. 
In this regard, the adversary may monitor the side-channel leakages to gather information, which can be further useful to predict the unseen bits generated by the TRNG, and/or mounting active attacks more effectively. 
As a prime example, passive temperature attacks can be considered~\citep{martin2014fault,cao2016exploring}, which significantly lowers the security level even of highly protected crypto-core implementations that are connected to the TRNG through reducing the entropy (i.e., introducing biased). 
The second category includes active attacks with the goal of disrupting the random number generation process, which are more prominent. 
To the best of our knowledge, one of the first active attacks has been reported in~\citep{vsimka2006active}, where by changing the temperature of the FPGA embodying a TRNG, the bias in its output is administered.

This line of research has further investigated frequency injection attacks against a security-dedicated Integrated Circuit (IC) embodying a two-ring oscillator (RO) based TRNG~\citep{markettos2009frequency}. 
It has been shown that when the power signal is manipulated by injecting a sine wave into that, the operating conditions of the two ROs can be modified, and consequently, the output of the TRNG is biased. 
This attack has been carried out on several different TRNGs working based on the jitter of the microcontroller's built-in RC oscillator~\citep{hlavavc2010true}, Self-Time Ring (STR)~\citep{cherkaoui2013very,cherkaoui2013self}, Transition Effect Ring Oscillator (TERO)~\citep{varchola2010new}, reported in~\citep{buchovecka2013frequency,martin2014fault,cao2016exploring}, respectively (see Table~\ref{tab:trng-attacks} for more examples). 

The research in this direction has been continued by studies, which investigate the impact of injecting electromagnetic (EM) signals into TRNGs. 
The effectiveness of such an attack has been recognized in~\citep{schmidt2007optical}, followed by studies enjoying the special features of this attack, namely being applicable through the package, and in particular, from the front-side cf.~\citep{ordas2015injection}. 
The authors of~\citep{bayon2012contactless} have launched such an attack on the RO-based TRNG. 

The main message to convey here is that on top of the vulnerability to the attacks mentioned above, TRNGs exhibit shortcomings such as a limited generation rate and being \emph{unpredictable}, rather than truly random.  
As mentioned earlier, QRNGs introduced to tackle some of these problems; however, the question remains open whether they can stand against attacks similar to one discussed in the section. 
The next two sections expand on these points.

\begin{table*}[t]
\caption{Summary of some of physical attacks on TRNGs}\label{tab:trng-attacks}
\scriptsize
\begin{center}
\begin{tabular}{|p{2.5cm}|p{4.5cm}|p{5.5cm}| } 
 \hline
Reference & Type of TRNG & Type of fault injected \\ \hline\hline
\citep{mesgarzadeh2005study}& ROs in general (not on TRNGs specifically) & Frequency locking of the oscillators\\\hline
\citep{bochard2010true} & RO-based & Manipulating the voltage supply \\\hline
\citep{soucarros2011influence}& RO-based & Temperature variation\\\hline
\citep{martin2014fault}& STR~\citep{cherkaoui2013very} & Clock glitches\\
\hline
\citep{madau2018impact}& Delay Chain True Random Number Generator~\citep{rozic2015highly} (RO-based)& Pulsed electromagnetic fault\\\hline
\citep{osuka2018information} & RO-based & Sinusoidal electromagnetic waves injection from a distance via power or communication cables attached to the target device embodying the TRNG\\ \hline
\citep{martin2018entropy}& RO-based~\citep{wold2008analysis} and STR~\citep{cherkaoui2013very}& Ionizing radiation\\ \hline
\citep{mahmoud2019timing}& STR~\citep{cherkaoui2013very} & Voltage drops induced remotely\\\hline
\end{tabular}
\end{center}
\end{table*}

\subsection{Defense: Quantum Random Number Generators}\label{sec:trng-qrng} 
Here we briefly introduce two main classes of randomness sources widely used for QRNGs, namely optical and non-optical sources\footnote{Note that this survey aims at providing an overview of QRNGs to give an understanding of their underlying mechanisms. We refer the reader to~\citep{stipvcevic2014true,herrero2017quantum} for further details. }. Moreover, we discuss certified QRNGs, and randomness extractors applied in the context of QRNGs, see Figure~\ref{fig:qrngs-taxonomy}. 

\subsubsection{Optical Randomness Sources for QRNGs} \label{sec:trng-qrng-opt-source}
Several parameters of the quantum states of light feature inherent randomness, and therefore, optical sources have been widely adopted to design QRNGs, see Table~\ref{tab:qrng-types}. 
These QRNGs are based on the premise that the quantum level of an optical field can be explained in terms of photons. 
Generating and detecting uncorrelated single photons is, therefore, the goal of various technologies, including single-photon detectors~\citep{jennewein2000fast,furst2010high,dixon2008gigahertz}, silicon detectors~\citep{ghioni2007progress}, superconducting nanowire single-photon detectors~\citep{marsili2013detecting,hadfield2009single}. 
Nonetheless, some of the shortcomings of single-photon detectors are their limited capabilities to count the photons, infeasibility due to the high cost, and long dead-time~\citep{herrero2017quantum}.   

The latter problem is not specific to single-photon detectors, but other optical QRNGs could also suffer from that, e.g., branching path-based generators equipped with two detectors located in different positions, where upon receiving a photon in any of these detectors, a bit can be generated cf.~ \citep{rarity1994quantum,jennewein2000fast}. 
For these systems, the long dead-time results in a low generation rate and high correlation among successive bits. 
On top of this, the mismatch between the detection efficiency of the detectors as well as the non-optimal coupling ratios of the beam splitter dividing the light between these detectors contribute to the bias in the generated bits. 
There are some proposals on how to overcome these shortcomings, in particular, integrated optical circuits inside silicon chips have been considered promising~\citep{grafe2014chip}. 
More concertedly, for these circuits, less variability and mismatch are expected; hence, the quality of the random bits could be improved. 

Besides the above optical QRNGs, a wide variety of systems and methodologies is used to quantum states of the light to generate random bits, e.g., QRNGs based on quantum vacuum fluctuation, photon counting, laser phase noise, to name a few. 
For more details on these QRNGs, we refer the reader to~\citep{herrero2017quantum}. 

\begin{table*}
\caption{Summary of proposals for QRNGs }\label{tab:qrng-types}
\begin{center}
\scriptsize

\begin{tabular}{|p{1.4cm}|p{1.8cm}|p{2cm}|p{2.5cm}|p{0.8cm}|p{3.5cm}| } 
 \hline
Randomness source & QRNG  & Representative reference & Detector & Rate & Weaknesses  \\
 \hline\hline
\multirow{5}*{Optical}& \multirow{4}*{Conventional}  &\citep{jennewein2000fast}  & Single-photon detector (SPD) & Mbps & Limited photon counting capabilities \\\cline{3-6}

  & & \citep{furst2010high}  & Photomultiplier tubes SPD & Mbps & Large variations in performance, limited efficiency, complex implementations\\\cline{3-6}
  
    & & \citep{ghioni2007progress,dixon2008gigahertz}  & Single photon avalanche photodiodes & Mbps & Limited performance\\\cline{3-6}
    
   & &\citep{marsili2013detecting} & superconducting nanowire SPD & NA & Low operating temperature, limited signal-to-noise ratio, higher probability of false detection events \\\cline{2-6}
   
   & Branching path &\citep{jennewein2000fast} & SPA & Mbps & Imperfect unbalanced detectors, detector dead time \\\cline{1-6}

\multirow{2}*{Non-optical}& Radioactive decay &\citep{lutz1999semiconductor} & PIN photodiodes & kbps & Complex implementation \\\cline{2-6}

 & Noise & \citep{stipvcevic2004fast} & comparator (an electronic circuit) & Mbps & Complex implementation\\\cline{1-6}
\end{tabular}
    \end{center}

\end{table*}

\subsubsection{Non-optical Randomness Sources for QRNGs} \label{sec:trng-qrng-nonopt-source}
Although optical randomness sources have been widely adopted in practice, proposals for non-optical QRNGs have been introduced, for instance, radioactive decay-based and noise-based QRNGs\footnote{For details on other classes of non-optical QRNGs, see~\citep{herrero2017quantum}. } (see Table~\ref{tab:qrng-types}). 
The former family is interesting due to the similarities between that and the optical QRNGs, whereas noise-based QRNGs are of great importance as commercial products have been developed in compliance with this principle, see, e.g.,~\citep{wilber2013entropy}. 

\vspace{0.5ex}

\noindent\textbf{Radioactive decay:} One of the first sources of randomness relying on quantum phenomena is radioactive decay. 
In this context $\beta$ radiation (emitted electrons) detectors have been mainly used, e.g., Geiger-M\"{u}ller cf.~\citep{friedman1949geiger}. 
In this regard, an analogy that can be drawn between these QRNGs and optical ones is interesting: 
the design of optical QRNGs is similar to ones applying the notion of radio decay, with the difference being that radioactive source and the GM counter are replaced by photon sources and detectors, respectively.

Generally speaking, radioactive decay-based QRNGs share some commonalities, namely using digital counters to convert the pulses from the detector into random digits, and digital clocks. 
Although being a high-quality randomness source, radioactive decay-based QRNGs suffer from low bit rate, being hard-to-access and measure, and inherent physical limitations (e.g., dead time -time to recover-, damage from radiations, etc.)~\citep{herrero2017quantum}. 

\vspace{0.5ex}

\noindent\textbf{Noise:} Noise as one of the most undesired characteristics of electronic circuits is, ironically enough, one of the most preferred sources of randomness in such circuits, see, e.g.,~\citep{stipvcevic2004fast}. 
For both TRNGs and QRNGs, a noise source can be coupled with a comparator to generate random bits. 
To this end, it can sometimes be necessary to amplify the noise before comparing that to a pre-defined threshold adjusted for the comparator. 
Furthermore, depending on how a random staring is formed, instead of sampling the output, a series of pulses can be generated by passing the signal from the noise source to the comparator.  
Regardless of this, when designing such a scheme, three main challenges to be faced are (1) proving the randomness of the noise source, (2) dealing with the effects of sampling/digitizing procedure, and (3) designing an adequate post-processing module to obtain close-to-ideal bias~\citep{stipvcevic2014true}. 

In order to design a noise-based QRNG, the noise corresponding to the shot fluctuation~\footnote{Shot noise can be produced from quantum effects due to the discrete nature of the electric charge and the particle nature of light in electrical and optical devices, respectively. } can be exploited; however, it is a challenging task to characterize and separate this type of noise from thermal noise~\citep{stipvcevic2014true,herrero2017quantum}. 
Even in some cases, the current value of the signal depends on the signal(s) passing through the circuit in the (near) past and can cause a correlation between random numbers generated, i.e., the so-called memory effect~\citep{stipvcevic2014true}. 

\subsubsection{Certified QRNGs}\label{sec:qrng-crtf}
Compared to TRNGs, for which it is challenging to guarantee the trustworthiness of either software or hardware, quantum mechanics enables us to assure this. 
More specifically, the quality of the output stream in terms of being unbiased and uncorrelated can be ensured through (1) self-testing, and (2) examining the inherent quantum statistical properties of a QRNG. 
In the first class, there are methods, whose underlying idea is comparable to embedded tests in the architecture of TRNGs, see Figure~\ref{fig:trng-blocks}. 
Nevertheless, these tests are primarily focused on measuring internal states and variables. 
As an example, to counteract the effect of maliciously controlling the quantum state that is the source of randomness, state tomography can be employed~\citep{james2005measurement}. 
However, for this type of test, repeated measurements on the same state should be granted. 
Furthermore, they are mainly effective against adversaries with limited access to the system, i.e., only the quantum state can be controlled. 
In addition to this, the security of such schemes depends heavily on the assumption that the QRNG works perfectly, i.e., without any memory effect cf.~\citep{acin2016certified}. 
Even for more sophisticated self-testing methods that can determine honest, technical noise and failures~\citep{lunghi2015self}, making the above assumption is necessary. 

\begin{figure*}[t]
    \centering
    \includegraphics[width=0.9\textwidth]{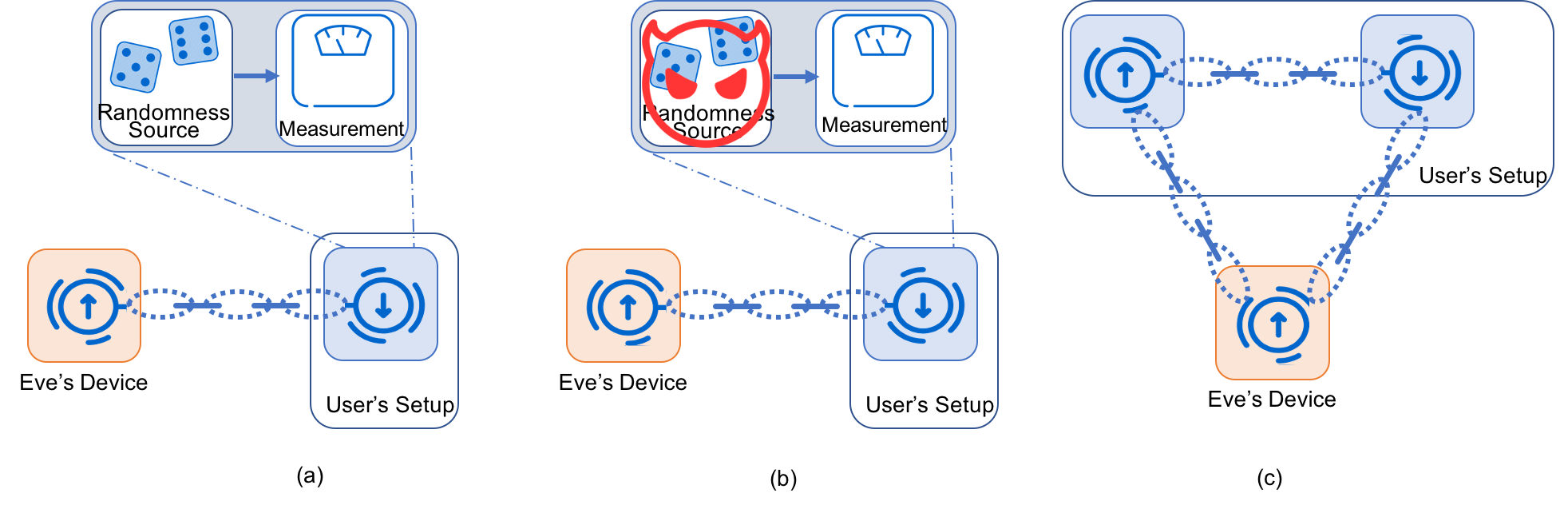}
    \caption{(Inspired by~\citep{acin2016certified}) 
    Some types of QRNGs discussed in Section~\ref{sec:trng-qrng}:
    (a) A typical QRNG employed by the user to generate bit strings that should be unpredictable by Eve, who can access any physical system other than the user's. Eve can also entangle her device with user's QRNG, where the entanglement is shown by the dashed chain.  
    (b) A semi-device-independent scheme, more specifically, source-device-independent QRNG. 
    In this case, the quantum state preparation can be imperfect, and even under the worst-case scenario, the device used to prepare the randomness is not trusted. 
    (c) A general schematic of the device-independent protocol. 
    The user's setup may contain more than two entangled devices. 
    Eve can also entangle her device with these devices. }
    \label{fig:qrngs}
\end{figure*}

\vspace{0.5ex}

\noindent\textbf{Device-independent QRNGs:} These schemes are another attempt to certify the random bits, where the randomness of the output is associated with physical laws, namely quantum non-locality, Bell's inequalities, and no-signaling principle, as explained below. 

Quantum non-locality states that for an entangled state, measurements performed independently on one of the separated particles set the state of the other particle. 
This correlation forms the basis of Bell's inequalities and quantum non-locality; that is, spatially separated systems have an influence on one another in a non-local manner, i.e., without any signal passing between them (see, Figure~\ref{fig:qrngs} and for a more detailed, precise definition, see~\citep{brunner2014bell}). 
This condition on the transmission of the signals is called the ``no-signaling principle'', which means that the information cannot be transmitted faster than the speed of light\footnote{Note that the correlation between the particles in an entangled state does not rule out the no-signaling principle. }. 

One of the first randomness generators relying on the Bell's inequalities has been suggested in~\citep{pironio2010random} (its improved version~\citep{pironio2013security}), where the randomness of the outputs measured from two separated systems is certified.   
In this regard, to generate an $n$-bit random string, a random seed should be given to the protocol, whose length is $\sqrt{n}$; hence, this protocol \emph{expands} the randomness. 
A similar method has been devised in~\citep{vazirani2012certifiable}, where the length of the seed could be reduced to $(\log-2n)^3$. 
It has been shown that the results presented in~\citep{pironio2013security,vazirani2012certifiable} can be extended to a more general model, where instead of quantum mechanics, solely the no-signaling principle must hold. 
Besides these, schemes designed to meet the no-signaling principle include mainly randomness amplifiers, see Section~\ref{sec:qrng-random-ext}. 

Another step towards reducing the reliance on quantum mechanics could be the study of randomness generators obeying more general no-signaling rules cf.~\citep{acin2016certified}. 
It has been shown that such theories studied in, e.g.,~\citep{barrett2005nonlocal}, lead to general non-signaling correlations; however, they cannot ensure the maximal randomness certification as the quantum theory does~\citep{barrett2006maximally}. 

In line with the above methodology, various certified QRNGs have been proposed in the literature, where other fundamental features of quantum theory are considered as an alternative for Bell's inequalities, namely contextuality~\citep{um2013experimental}. 
The main advantage of this type of certification is that randomness from quantum origin can be differentiated from the randomness due to classical noise and imperfections or failures in the QRNG cf.~\citep{herrero2017quantum}. 
Moreover, it is not required to ensure space-like separation between two systems involved in the protocols, e.g., as applied in~\citep{bierhorst2018experimentally}. 
Nonetheless, in contrast to device-independent QRNGs discussed above, the security of such systems cannot be guaranteed against malicious manufacture. 

\vspace{0.5ex}

\noindent\textbf{Semi-device-independent QRNGs:} 
The core idea behind these QRNGs is extending the concept of device-independent QRNGs beyond the notion of Bell's inequalities, see, e.g.,~\citep{li2012semi}. 
In this case, to certify quantum randomness, an experimental setup composed of a preparing device and a measuring device. 
Whereas the measuring one performs measurements, the former device is employed to prepare a system in different quantum states, assumed to be in a Hilbert space with a bounded dimension. 
The certification is performed by examining dimension witnesses~\citep{gallego2010device}. 
Although this allows non-entangled devices to be used, the efficiency of the protocol should be verified experimentally. 
This shortcoming has been tackled in~\citep{bowles2014certifying} by showing that if (a) the preparation and measuring device share no correlations, and (b) devices do not suffer from memory effects, the randomness is certified regardless of the qubit detection efficiency. 
Clearly, the second assumption cannot hold easily in practice, and therefore, the security of a practical system is not proved. 

In spite of the above disadvantage, a series of work has reported how semi-device-independent QRNGs can be designed with regard to a set of assumptions~\citep{vallone2014quantum,van2017semi}. 
For instance, an assumption made regarding the preparation device, that is, the amount of true randomness, which can be obtained by a given source, can be estimated~\citep{vallone2014quantum}. 
As another example of these, so-called, source device-independent schemes,~\citep{smith2019simple} proposes a QRNGs with the goal of enhancing the practical implementation of semi-device-independent devices. 
To this end, no additional optical components are required to implement their continuous-variable optical quantum random number generator. 

\subsubsection{Randomness Extractors}\label{sec:qrng-random-ext}
For TRNGs, the idea behind the design of random extractors is to cancel out the impact of aging, temperature variations, and even, attacks. 
First designed for this purpose~\citep{barak2003true}, extractors have found application for QRNGs. 
The fact that true randomness is a key, intrinsic feature of quantum mechanics employed in the context of QRNGs does not preclude the need for randomness extractors. 
Nevertheless, the main goals of these schemes are quantum randomness expansion and quantum randomness amplification~\citep{herrero2017quantum}. 

Similar to seeded randomness extractors, methods in the first category aim at obtaining a relatively long, random bit sequence by feeding a small random seed to a quantum protocol. 
This can be fulfilled through the composition, i.e., by concatenating a finite number of quantum devices, which has been shown to be secure against quantum adversaries, see, e.g.,~\citep{coudron2014infinite,miller2017universal}. 
It is also possible to go one step beyond this: given imperfect quantum sources, a randomness extractor can be employed to certify the randomness of the output~\citep{cao2016source}. 
In doing so, one can address shortcomings of practical QRNGs, including losses, multiphoton pulses, or unbalanced beam splitters affecting optical QRNGs. 
In addition to these, in a broad sense, some of the device-independent QRNGs can fall into this category since they also employ weak randomness generated in the non-locality experiments of Bell's inequalities. 

Clearly, for the randomness expansion, a uniform seed is required to either run the Bell test (i.e., choosing the measurement setting) or improve the entropy for achieving a uniform output bit string. 
To eliminate this requirement, yet meet the needs mentioned above, quantum randomness \emph{amplification} methods have been proposed. 
In their design, a weak source is combined with \emph{independent} quantum devices, where these devices are often restricted to simple measurements on the different subsystems of an entangled state~\citep{herrero2017quantum}. 
In this regard, one of the most celebrated methods has been proposed in~\citep{colbeck2012free}; however, a large amount of imperfect randomness is required. 
This has been addressed by Ramanathan et al.~\citep{ramanathan2016randomness} that demonstrate a protocol requiring two devices. 

Furthermore, in~\citep{colbeck2012free}, conditions on the min-entropy of a weak source given as an input to the protocol have been defined, which can be hard to fulfill in practice. 
As a remedy for this difficulty, approaches presented in~\citep{bouda2014device,chung2014physical,plesch2014device} can be considered. 
To further improve the robustness of random amplifiers against noise, a protocol has been suggested in~\citep{brandao2016realistic}, where using a finite number of devices, arbitrary weak randomness can be amplified into nearly perfect random bits. 

\begin{figure*}[t]
    \centering
    \includegraphics[width=\textwidth]{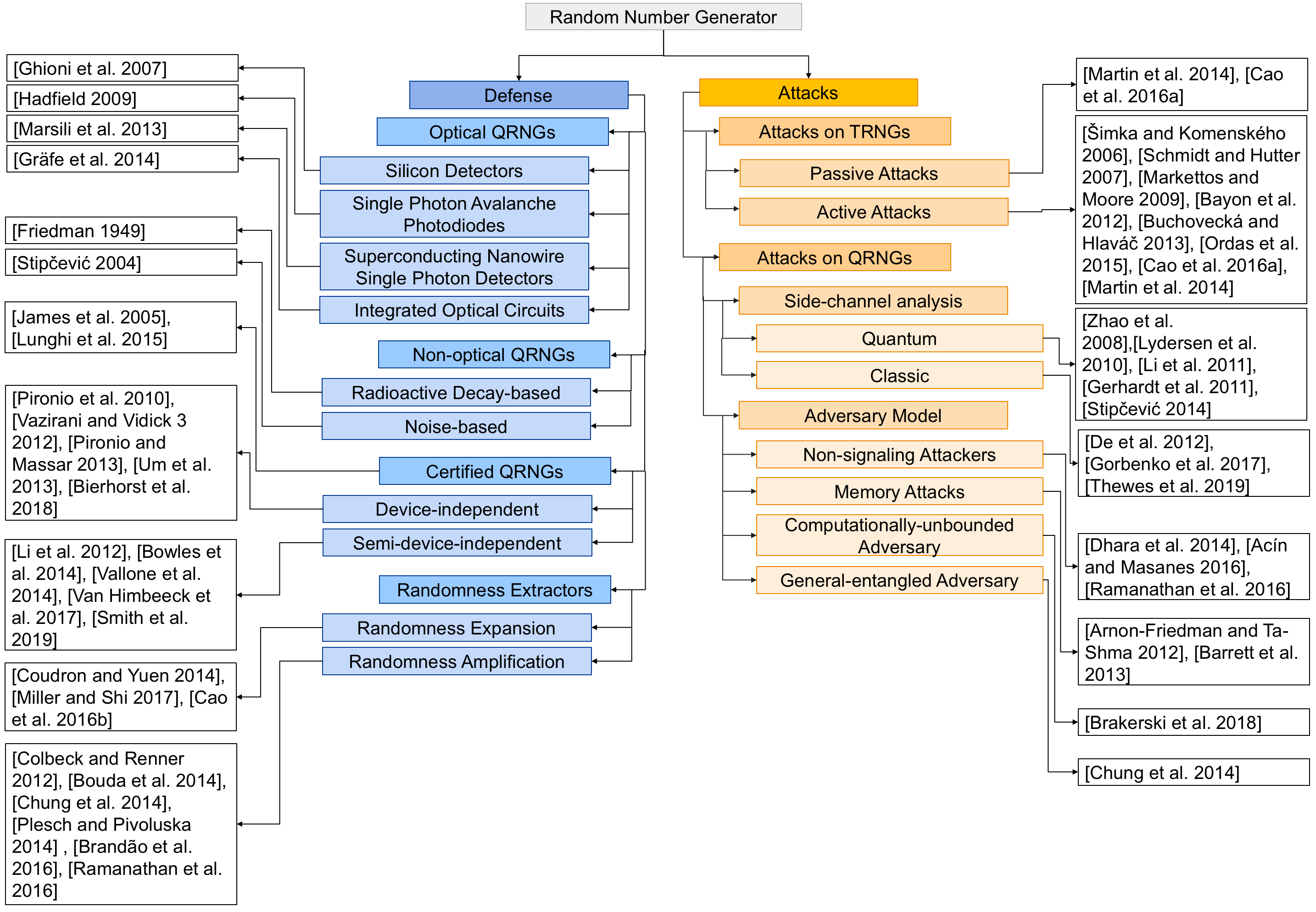}
    \caption{Taxonomy of methods devised to offer randomness in the quantum world. Attacks and adversary models discussed to assess the security of TRNGs and QRNGs have been depicted as well. 
    Note that only some of references reviewed in this paper are mentioned here. }
    \label{fig:qrngs-taxonomy}
\end{figure*}

\subsection{Attacks and Adversary Models}
Among the premises, on which QRNGs are based, are providing trusted randomness sources and generating numbers at a faster rate. 
While the latter can be required for specific purposes, e.g., online services relying on a stream of random data, the former requirement is imposed by virtually all applications. 
In contrast to TRNGs, where the source of the randomness should be examined to be random or, at least, exhibit an acceptable level of unpredictability, QRNGs harvest well-defined, intrinsically random processes. 
Clearly, this feature of QRNGs makes them especially suitable for quantum cryptography, with the need for a reliable randomness source. 

On the other hand, similar to TRNGs, QRNGs can become an Achilles heel of cryptographic systems. 
For instance, attacks against QKDs mainly target QRNGs (or randomness generator modules) as the heart of these systems. 
In these attacks, the entire key may be extracted by the adversary seamlessly, i.e., the honest parties exchanging the keys cannot find out that an attack has happened. 
This is indeed against conditions defined for and features expected from random number generators, namely unpredictability, and backward security. 
While unpredictability ensures that an attacker knowing the whole sequence cannot guess the next bit with a probability better than one-half, backward security means that knowing a part of the sequence does not result in the disclosure of previous values of the generator with better accuracy than guessing~\citep{herrero2017quantum}. 

This section explores attacks mounted on QRNGs that, in contrast to the above requirement, can break the security of QRNGs. 
Moreover, adversary models suggested in the literature to evaluate the security of QRNGs are discussed, see, Figure~\ref{fig:qrngs-taxonomy}. 

\subsubsection{Side-channel Analysis} 
As explained in Section~\ref{sec:trngs-design-attacks}, attacks targeting TRNGs essentially take advantage of the information leaked from TRNGs by either measuring the side-channels or by injecting a fault and then, measuring the side-channels. 
Similar to theses classical counterparts, attackers can perform side-channel analysis on quantum devices.  
Although such attacks share several similarities with ones launched on TRNGs, there are some differences between them: (a) to conduct side-channel analysis on QRNGs, access to the device(s) may be required, (b) the adversary can entangle her device with the devices used by honest parties to extract quantum side-channels. 
One can, of course, draw an analogy between the latter difference and the fault injection attacks launched on TRNGs. 

First attempts at launching side-channel attacks on quantum devices have been made to break the security of QKD systems. 
In this respect, e.g., by utilizing possible correlations between different degrees of freedom of the photons (spatial, spectral, or temporal properties) with the actual bit values, exchanged information can be revealed. 
Note that conducting such an analysis does not lead to an error. 
As a prime example, the time-variant detection efficiency of photon detectors has been exploited to launch an attack resulting in a bias~\citep{zhao2008quantum}.  

Side-channel information leakage caused by arbitrary imperfections in the implementation of a QRNG forms another aspect of side-channel attacks. 
Timeshift attacks can fall within this category, where the mismatch between the detection efficiency of detectors for the bits ``0'' and ``1'' is exploited~\citep{zhao2008quantum}. 
In this regard, the arrival time of each signal is manipulated by the adversary to induce bias in the system. 
Besides these attacks, on the same principles, detector blinding attacks reported in~\citep{lydersen2010hacking,gerhardt2011full} have been proposed cf.~\citep{stipvcevic2014preventing}. 
For these attacks, it is crucial that the adversary can control the detectors at the receiver side by performing, e.g., strong illumination that eliminates the randomness in the measurement. 
In this context, attacks enjoying the undesirable, yet inevitable characteristics of beam splitters, that is, wavelength-dependency have been studied in the literature~\citep{li2011attacking}. 

Another type of side-channel investigated in the literature can be referred to as ``classical'' side-channel, where the quantum-related features of the system are \emph{not} employed, but non-trivial information about, e.g., the seed. 
The security of randomness expansion and recycling schemes can be studied under this scenario, see, e.g.,~\citep{impagliazzo1989recycle,de2012trevisan}. 
A more interesting case has been examined in~\citep{thewes2019eavesdropping}, where classical side-channels, namely the classical noise, can provide the attacker with information about the random number generated by honest parties. 

Recently, a new study has been carried out to explore the possibility of mounting the EM side-channel attack on a QRNG~\citep{gorbenko2017differential}. 
For this attack, it has been claimed that the EM signal leaking from the extractor can be used to break the security of a QRNG. 
Although the idea may seem plausible, further research on this is required to investigate various aspects of this attack and its feasibility. 

\subsubsection{Adversary Models}
When it comes to evaluating the randomness of a random number generator, the power of an adversary constitutes the other side of the coin. 
More specifically, we are interested in examining if the output of the generator is random from the point of view of an attacker with a specific power. 
Below, we describe models that take into consideration different power levels for an adversary. 

\vspace{0.5ex}

\noindent\textbf{Non-signaling attackers:}
This model has been widely adopted in the literature, in particular, studies on randomness amplification, see, e.g.,~\citep{ramanathan2016randomness,acin2016certified,dhara2014can}. 
In this model, there is no restriction imposed on the adversary rather than the non-signaling principle. 
This principle states that faster-than-light communication between devices, regardless of belonging to honest or malicious parties, is impossible. 

Note that although when considering this type of adversaries, it is not needed to rely on the validity of quantum mechanics, QRNGs are still expected to generate correlations explained by Bell's inequalities. 

\vspace{0.5ex}

\noindent\textbf{Memory attacks:} 
The scenario described by this model represents an exceptional situation, where the device manufacturer misuses her access to the QRNG to launch the following attack~\citep{barrett2013memory}. 
The QRNGs is run to generate a long bit string, which is then stored into a memory stick and sold as a proper generator to the user; however, the adversary (i.e., malicious manufacturer) can perfectly predict the bits in the string. 
Needless to say that the bit string provided to the users passes any statistical test and looks random. 

Memory attacks have also been discussed under another scenario, where the memory embedded in a device could be exploited to make information from past measurements available for future measurements. 
Obviously, this is against the notion of the non-signaling principle; however, the goal of the study presented in~\citep{arnon2012limits} is to demonstrate that in the presence of such memory, randomness amplification can be impossible. 

\vspace{0.5ex}

\noindent\textbf{Computationally unbounded adversary:} The authors of~\citep{brakerski2018cryptographic} have studied the design of an interesting certifiable QRNG, where instead of multiple quantum devices sharing entanglement, a single computationally-bounded quantum device has been considered to run the proposed protocol. 
This has been certified through the reduction to the hardness of the learning with errors (LWE) problem. 
It has been further shown that even for a computationally \emph{unbounded} adversary, who may share entanglement with the generator, the output remains statistically indistinguishable from a uniformly random bit string. 
In fact, this type of adversaries is extremely powerful; however, it has been proven that they cannot leverage the information leaked from the entangled devices. 

\vspace{0.5ex}

\noindent\textbf{General entangled adversary:} 
This model has been introduced to assess the security of extractors that harvest the randomness from multiple independent sources.  
In this model, as its name implies, the adversary with access to quantum sources share entanglement with the sources to obtain side-channel information~\citep{chung2014physical}. 
In order to remove interference among the sources, when measuring the side-channel, it is suggested to conduct the measurement on each and every source individually.
This model can be considered as a generalization of models studied in~\citep{kasher2012two} in the sense that limits have been imposed on neither how the side-channels from different sources are combined nor the amount of information stored by the adversary cf.~\citep{arnon2015quantum}. 
The latter condition is known as the ``bounded-storage'' model, widely applied in work related to the design of randomness extractors, see, e.g.,~\citep{ta2011short,de2012trevisan}. }


\section{Physically  Unclonable Functions}\label{sec:pufs}
\sloppypar{

Physically unclonable functions (PUFs) cater to the requirement of a unique, unbiased yet reliable fingerprint for anti-counterfeiting, responses for authentication, and volatile key for cryptographic schemes in today's world. PUFs are used for device authentication and key-generation purposes for cryptographic algorithms and form the backbone of major authentication protocols used in cryptography. PUFs are also popular for secured storage of sensitive data like cryptographic keys and are one of the major research topics among security primitives. The initial idea of such unclonable, repeatable, and volatile secrets was first coined as ``physical one-way functions'' by R. Pappu et al. in \citep{Pappu-Science}. Later, it was extended to silicon by utilizing the process variations of a chip by Gassend et al. in \citep{Gassend-2002} for ``silicon PUFs''. Silicon PUFs utilize the random manufacturing variations within a silicon chip and generate a unique, unclonable identification key from it. The output of a PUF is recorded in the form of challenge-response pairs (CRPs) which, are generated by physically querying the PUF and computing its response. Research in silicon PUFs has investigated secret generation based on metastability, race conditions, etc. in different units present in a silicon chip including memory (DRAM, SRAM, etc.), ring oscillators (ROs), comparators, latches, current mirrors, etc. 

In the literature, PUFs have been categorized based on their structure, performance, or the number of CRPs generated by them. In this paper, we will separate PUFs into two categories: PUFs in the classical era and PUFs in the post-quantum era. According to \citep{arapinis2019quantum}, classical PUFs generate classical CRPs and also limit the adversary to only classical interactions with the PUF. In other words, both the authentication and key-generation process include physical characteristics that are only defined by classical mechanics and do not adhere to generation or interaction with quantum particles or quantum bits. With the advent of the state-of-art research on quantum supercomputers as described in Section \ref{sec:qcrypto}, restricting the adversary to only the classical domain may camouflage potential vulnerabilities within the PUF architecture, thus leading to incorrect evaluation. So, for correct performance and security evaluation, PUFs must also be assessed in the quantum domain where the adversary is allowed to perform quantum interactions with the PUF system. This motivates us towards a new perspective of PUFs in the post-quantum era. Overall, we have divided the state-of-art research for PUFs in the quantum domain into two major divisions, a) \textit{Defense} covers the architectures and design of several PUFs as well as authentication protocols with both quantum and classical architectures, built to defend against cloning, piracy, and attacks incorporating quantum attacks. This can be seen in the taxonomy shown below, as depicted in  Figure~\ref{fig:PUF-taxonomy}; b) \textit{Attacks} against PUFs in both classical and quantum domains. We describe both the domains in detail in the following sections providing perspectives for current state-of-art research in PUFs.

\subsection{PUF Preliminaries: Pre-Quantum Era}
Silicon PUFs utilize the process variations within a chip to generate unique fingerprints. Thus, two chips, in spite of possessing the same design and architecture, will produce different CRPs, leading to a one-to-one mapping of every individual chip produced by a company. State-of-the-art work on silicon PUFs can be divided into three types as shown in~\citep{Zhang2014}: 1) \textit{Delay based PUFs} consist of a delay chain producing unique delays due to process variations as seen in Arbiter PUF, RO PUF,  etc. \citep{Suh-2007}; 2) \textit{Memory based PUFs} are mainly comprised of memory cells like SRAM~\citep{SRAM-patent}, DRAM~\citep{Tang-CICC-2017}, etc. and use their start-up characteristics to generate unique binary strings. 3) \textit{Analog electronic PUFs} form a rather contemporary category targeted towards analog and mixed-signal ICs. Some instances include mono-stable current mirror PUF in~\citep{Alvarez-ISSCC}, PTAT (proportional to absolute temperature) PUF in~\citep{Li-ISSCC}, etc.
It must be noted that these existing PUFs are meant completely for the classic world and do not have any quantum connotations associated with them. Neither the architecture is based on quantum phenomena, nor the architecture provides any robustness against quantum attacks. Thus, loosely these PUFs can be termed as PUFs in the pre-quantum era.
The most commonly used metrics for PUFs aim to measure their suitability for the aforementioned applications. A PUF signature should be impossible to recreate. In other words, there must be a large variation between the responses of any two PUF instances (in different devices). This referred to as ``uniqueness'' of a PUF and is measured by the inter-chip Hamming Distance (HD). The ``reliability'' of a PUF, i.e., ability to recreate its output upon measurement, is measured by intra-chip HD.

Classical PUFs belong to a mature research field that includes various architectures and authentication schemes. Despite several research works that have been conducted in the classical PUF domain, many of them can be vulnerable to attacks in the post-quantum era. In our future sections, we describe in detail the PUFs which belong to quantum-era and the advantages/disadvantages associated with it.

\subsection{Defense: Innovative PUF architectures for Anti-counterfeiting and for Preventing Quantum Attacks}\label{subsec:pufs-defense}

 While reviewing PUFs in the post-quantum era, there are two interdependent aspects that one needs to consider, as shown in the taxonomy in Figure~\ref{fig:PUF-taxonomy}, a) \textit{Quantum-generated PUFs or Quantum PUFs} that utilize properties inherent in quantum interactions for PUFs. Note that these PUFs may or may not be secure against quantum adversaries, which takes us to b) \textit{Classical quantum-secure PUFs}. These represent examples of PUFs that have already been proven secure against quantum attacks. A detailed representation of the characteristics of the quantum generated and quantum secure PUFs have been presented in table \ref{tab:QG-QS-PUF}. 
 It must be understood that the existence of a quantum adversary can jeopardize the security of PUFs that have already been claimed secure against traditional machine learning (ML) attacks. Thus, we require special architecture for PUFs that provide resilience against quantum attacks. In conclusion, the availability of quantum systems can completely transform the traditional viewpoint for PUFs. In our next subsections, we try to answer the persisting questions regarding PUFs in the post-quantum world, imploring a new line of research approach for PUFs.
 
 \subsubsection{Quantum-generated PUFs or Quantum PUFs (QPUFs)}
 In contrary to classical PUFs, quantum-generated PUFs originate owing to a certain quantum phenomenon associated with any device. 
 The inherent properties of a qubit can bolster the architecture of a PUF in meaningful ways. A qubit is immediately lost after it has been measured; thus, it is impossible to physically clone a qubit. If the adversary is unaware of the measurement procedure, it is almost impossible for an adversary to reproduce a qubit. Thus, quantum properties of a device can be utilized to generate a PUF, and a few such instances have appeared so far in literature for novel quantum PUFs.
 \begin{figure*}
    \centering
    \includegraphics[width=1\textwidth,height=1\textwidth,keepaspectratio]{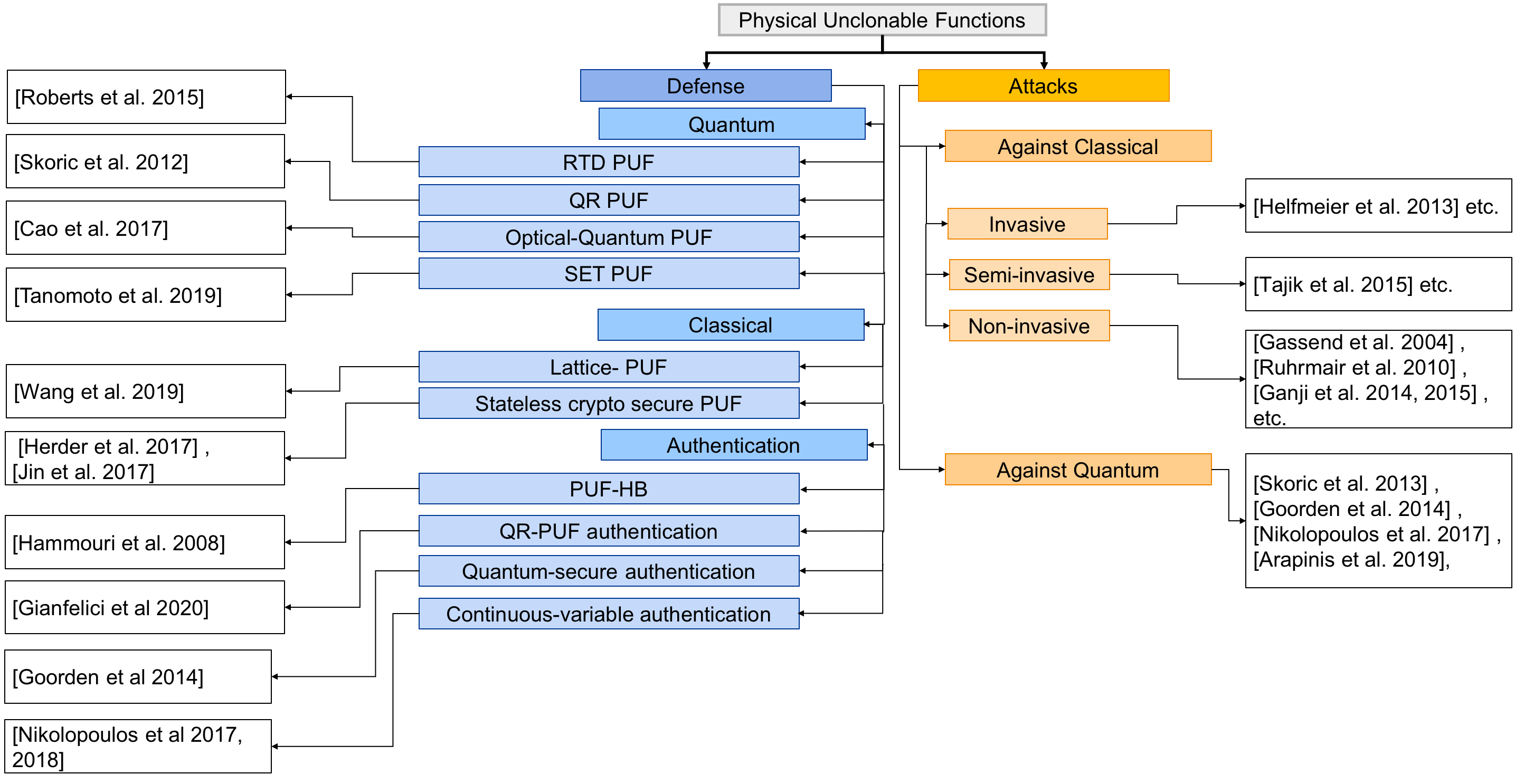}
    \caption{PUF Taxonomy in post-quantum era showing research work explored in both defense and attack categories.}
    \label{fig:PUF-taxonomy}
    \captionsetup{belowskip=0pt}
    \vspace{-1\baselineskip}
\end{figure*}

\vspace{0.5ex}

\noindent\textbf{Resonant Tunneling Diode (RTD) - PUF:} In \citep{Roberts-nature-SR}, the authors use the quantum tunneling using a resonant tunneling diode (RTD) consisting of a quantum well to generate a PUF. RTD consists of a quantum well surrounded by two energy barriers such that only electrons possessing a specific energy level can tunnel through the device. The voltage across the device represents the ratio of the energy of the confined quantum well to the energy level of the emitter, whereas the current through the diode indicates the total number of electrons tunneling. The authors inspect the current-voltage curves generated from each RTD device. 
The position of the current peaks with respect to the applied voltage presents an inherent quantum property of each device, which can be used to generate a quantum PUF. Due to process variation, the position of the current peak is unique for a single device, thus generating a fingerprint depending on the quantum property of the device. The authors explain that such a quantum PUF is capable of producing approximately 1000 unclonable identifiers, which can be further increased by putting multiple devices in an array. Since the authors have not mentioned anything about the reliability of the RTD PUF, it is arguable how much the device signature is repeatable across environmental and temporal variations. 

\vspace{0.5ex}
\begin{center}
\scriptsize
\begin{table*}
\centering
\caption{Quantum-generated and quantum-secure PUFs and their respective characteristics}
\scriptsize

\begin{tabular}{|p{1.1cm}|p{2.7cm}|p{1.4cm}|p{1.6cm}|p{2.1cm}|p{3.3cm}| p{2.3cm}|} 
\hline
Type & Name & \begin{tabular}[c]{@{}l@{}}Phenomena \\ used\end{tabular} & Challenges & Response& \begin{tabular}[c]{@{}l@{}}Security \\ advantages\end{tabular} & Applications \\ \hline
\multirow{4}{*}{\begin{tabular}[c]{@{}l@{}}Quantum\\ generated\\  PUFs\end{tabular}} & \begin{tabular}[c]{@{}l@{}}RTD PUF \\ \citep{Roberts-nature-SR}\end{tabular} & \begin{tabular}[c]{@{}l@{}}Quantum \\ tunneling\end{tabular} & \begin{tabular}[c]{@{}l@{}}Voltage \\ across \\ device\end{tabular} & Current peak & \begin{tabular}[c]{@{}l@{}}Unclonability, uniqueness\\ with minimal resources \\ \& small dimension\end{tabular} & \begin{tabular}[c]{@{}l@{}}Security application\\ requiring small area,\\ width, power etc.\end{tabular} \\ \cline{2-7} 
 & \begin{tabular}[c]{@{}l@{}}QR PUF \\ \citep{Skoric-2012}\end{tabular} & \begin{tabular}[c]{@{}l@{}}Optical \\ scattering\end{tabular} & \begin{tabular}[c]{@{}l@{}}Single \\ photons\end{tabular} & Quantum states & \begin{tabular}[c]{@{}l@{}}Remote authentication\\ without trusted readers\end{tabular} & \begin{tabular}[c]{@{}l@{}}Authentication\\ protocols\end{tabular} \\ \cline{2-7} 
 & \begin{tabular}[c]{@{}l@{}}Optical Quantum  PUF \\ \citep{Cao-2017}\end{tabular} & \begin{tabular}[c]{@{}l@{}}Defects in\\ mono-layer\\ materials\end{tabular} & \begin{tabular}[c]{@{}l@{}}Incident light\\frequency \&\\ intensity\end{tabular} & \begin{tabular}[c]{@{}l@{}}Output \\ multi-frequency,\\ multi-intensity\\ spectrum\end{tabular} & \begin{tabular}[c]{@{}l@{}}Unclonability, uniqueness\end{tabular} & \begin{tabular}[c]{@{}l@{}}Counterfeit \\ detection\end{tabular} \\ \cline{2-7} 
 & \begin{tabular}[c]{@{}l@{}}SET PUF \\ \citep{Tanamoto-APL}\end{tabular} & \begin{tabular}[c]{@{}l@{}}Single \\ electron\\ tunneling\end{tabular} & Gate voltage & \begin{tabular}[c]{@{}l@{}}Drain current\\ generating\\ patterns called \\ coulomb diamonds\end{tabular} & \begin{tabular}[c]{@{}l@{}}Unclonability, uniqueness\end{tabular} & \begin{tabular}[c]{@{}l@{}}Device \\ authentication\end{tabular} \\ \hline
\multirow{2}{*}{\begin{tabular}[c]{@{}l@{}}Quantum\\ secure \\ PUFs\end{tabular}} & \begin{tabular}[c]{@{}l@{}}Lattice PUF\\ \citep{wang2019lattice}\end{tabular} & \begin{tabular}[c]{@{}l@{}}SRAM POK\\ \& LWE \end{tabular} & \begin{tabular}[c]{@{}l@{}}Classical \\ digital\\ challenges\end{tabular} & \begin{tabular}[c]{@{}l@{}}Classical SRAM\\ startup response\end{tabular} & \begin{tabular}[c]{@{}l@{}}Not PAClearnable; secure\\ against classical \& quantum\\ ML attacks\end{tabular} & \begin{tabular}[c]{@{}l@{}}Authentication\\ requiring security\\ against classical \&\\ quantum attacks\end{tabular} \\ \cline{2-7} 
 & \begin{tabular}[c]{@{}l@{}}Stateless PUF\\ \citep{Trapdoor-herder}\end{tabular} & Hardness of LPN & \begin{tabular}[c]{@{}l@{}}Classical\\ digital\\ challenges\end{tabular} & \begin{tabular}[c]{@{}l@{}}Difference in RO\\ frequncy\end{tabular} & \begin{tabular}[c]{@{}l@{}}LPN-hard; secured against\\ classical \& quantum \\ attacks\end{tabular} & \begin{tabular}[c]{@{}l@{}}Authentication \\ requiring security\\ against quantum \\ attacks\end{tabular} \\ \hline
\end{tabular}
\label{tab:QG-QS-PUF}
\end{table*}
\end{center}

\noindent\textbf{Quantum readout (QR) - PUF:} In \citep{Skoric-2012}, the authors developed a quantum-readout (QR) PUF, where a classical PUF was challenged by a quantum state (referred as a single photon-state) generating a quantum response. For example, an optical PUF, challenged by photon states instead of a classical light beam producing a quantum state as a response. In this case, they used single-photon states through an optical fiber to challenge the PUF. The number of challenges was limited owing to the sparse number of transversal modes carried by the optical fiber. Yet, the presence of multiple wavelengths aided the increment of the number of available challenges. The major advantage of the QR-PUF, as discussed by the authors, is the possibility of remote authentication without the availability of trusted readers. The remote authentication protocol has been further extended in \citep{Skoric2017}, which allows the receiver to verify that the data has been sent by the PUF holder. In short, the unclonability of quantum states prevents the attacker from intercepting the data sent over by the QR-PUF. In case of an interception, the PUF response changes owing to the inherent property of quantum states, warning the receiver. It has also been assumed that all the properties of QR-PUF are public except the challenge states, and with the help of security analysis, the authors prove that it is not feasible to build a quantum computer that can emulate the QR-PUF. 

\vspace{0.5ex}

\noindent\textbf{Optical Quantum PUF}: An optical quantum PUF as described in \citep{Cao-2017}, uses nanoscale defects within 2D sheets to uniquely identify devices. Imperfections within monolayer materials like transition metal dichalcogenides (TMD) alter the bandgap structure of the semiconductor. 
Thus, when light is incident on the monolayer, it generates a spectrum of lights with various frequencies and intensities at different locations, depending on the intrinsic imperfections. The emitted spectrum is collected by a lens and passed through a band-pass filter (BPF). 
The angular orientation and the bandwidth of the BPF, as well as the spatial position across the monolayer, can be seen as the challenge to the PUF whereas the intensity and frequency of the light emitted by the monolayer and later collected by a charge-coupled device (CCD) can be seen as the response. Unclonability and utilization in counterfeit detection have been claimed as the major usage for this PUF by the authors.

\vspace{0.5ex}

\noindent\textbf{PUF based on single-electron tunneling (SET)}: Another instance of a quantum PUF surfaced recently in \citep{Tanamoto-APL}, which, can generate device fingerprints using the SET effect. When a countable number of electrons is confined within a small space then, SET can be observed. SET generates quantum dots (QDs) containing confined electrons at discrete energy levels, which produces unique fingerprints in measured drain currents relative to the varying gate voltage. The technology can be used to identify chips and represent a "quantum version of PUF,"  as described by the authors. 

The inherent no-cloning property of quantum gates and algorithms can prove immensely beneficial for security primitive architectures and is one of the main advantages of the above PUFs designed from quantum phenomena. Nonetheless, it is still a beginning for quantum-generated PUFs and thus requires fresh ideas and research in the domain. An important research direction can be the quantification of metrics for quantum PUFs. The metrics for classical PUFs have been well-researched, but it must be noted that similar metrics may not be sufficient for evaluating quantum PUFs since there are differences in their basic principles of operation. The major metrics signifying performance for a classical PUF include uniqueness, reliability, and randomness. While most of the above quantum PUFs provide a unique output, reliability is still a question for them. According to our knowledge, very little has been specified about the reliability of these quantum PUFs. Also, to calculate reliability for quantum PUFs, it must be understood whether the quantum properties change at all with varying external conditions. More specifically, the type of conditions against which reliability needs to be tested must be formulated. It has been shown in \citep{Giulio-PRA} that certain fundamental properties of quantum PUFs can adversely affect the reliability. Here, the authors mention that the property of non-orthogonality of the quantum challenges provide unclonability but also can introduce errors in measurement, thus reducing the reliability.

 \begin{figure*}
    \centering
    \includegraphics[width=0.8\textwidth,keepaspectratio]{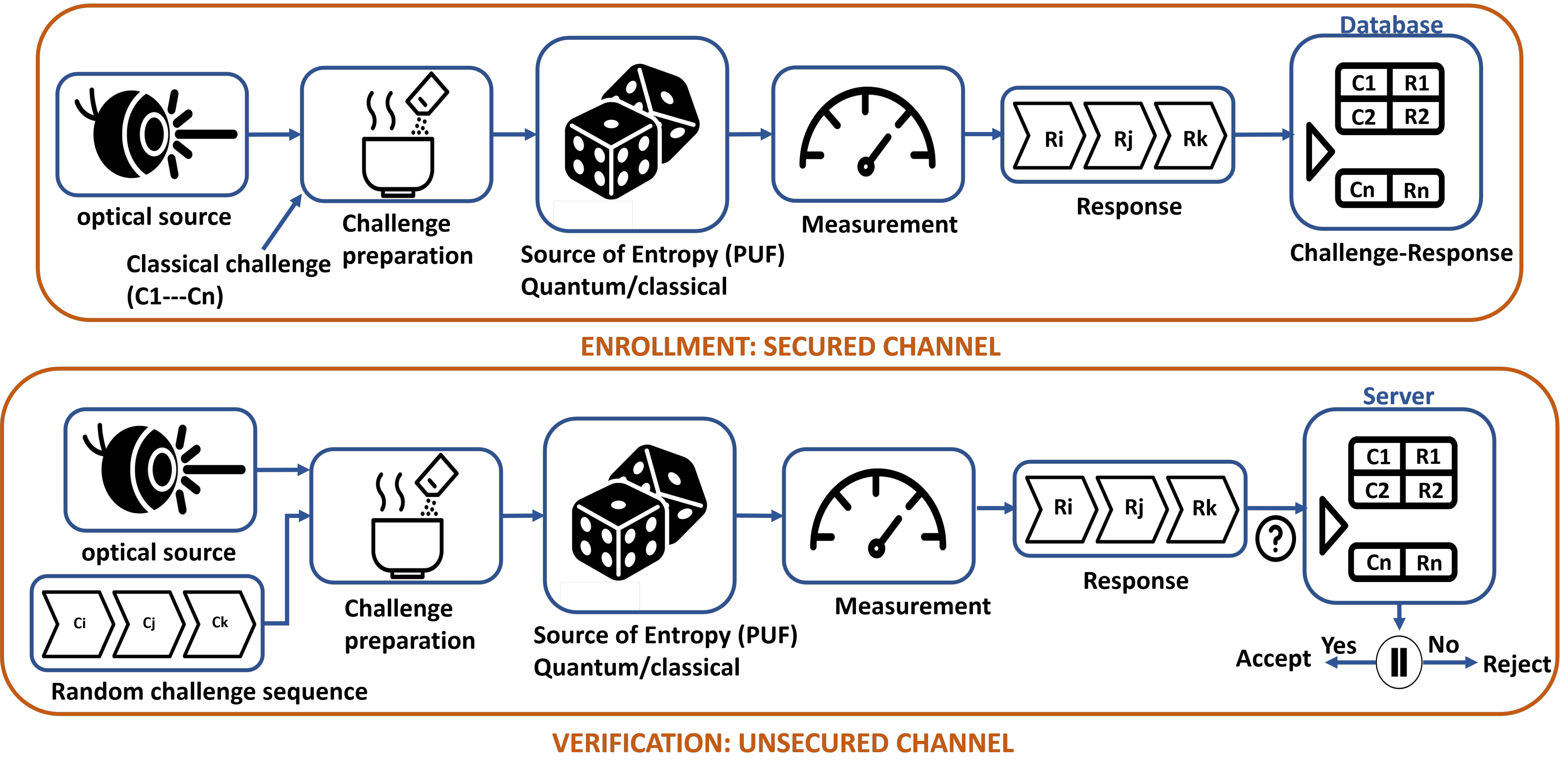}
    \caption{A general structure representing the authentication protocol using PUFs consisting of enrollment and verification stages.}
    \label{fig:PUF-EAP}
    \captionsetup{belowskip=0pt}
    \vspace{-1\baselineskip}
\end{figure*}
\subsubsection{Classical Quantum-secure PUFs}
The advent of quantum technologies has increased the vulnerabilities of security primitives, assisting attacks against PUFs by reducing the complexity of computation and time to brute-force. Thus designing PUFs that are provably secure against quantum attacks remains a priority. There are a few instances of classical PUF architectures that have been proposed so far, which provide resistance against ML attacks in both classical as well as quantum domains. It must be noted that the design architecture of all these PUFs pertain to classical systems and do not generate responses based on quantum interactions. Nevertheless, the responses provided by these PUFs can be mathematically proven to be secure against quantum attacks based on their inherent architectures.  

\vspace{0.5ex}

\noindent\textbf{Lattice PUF}: In \citep{wang2019lattice}, a strong PUF is built, which is claimed to be provably secure against ML attacks both in classical and quantum computations. The lattice PUF comprises of a physically obfuscated key (POK) and a learning with errors (LWE) decryption function that generates the response relative to the challenges given to the PUF. To evaluate the robustness of the lattice PUF, the authors introduce a metric called ML resistance. A PUF is defined to have $k$-bit ML resistance if an ML attack requires $2^k$ attempts to break it. The proposed lattice PUF produces a CRP size of $2^{136}$ and can provide 128-bit ML resistance. The authors claim that the major advantage of the lattice PUF is that it derives its security from the fact that the decryption function used in the lattice PUF is not PAC (probably approximately correct framework)-learnable.

\vspace{0.5ex}

\noindent\textbf{Stateless Cryptographic PUF}: In \citep{Trapdoor-herder}, a cryptographically secure PUF is built from learning parity with noise (LPN) hard fuzzy extractor. Fuzzy extractors are cryptographic structures that eliminate the noise from biometric data (either human or silicon) and generate random, uniform keys for security and cryptographic applications. The authors at first provide a fuzzy extractor that can correct $O(m)$ errors in polynomial time from a biometric source. This fuzzy extractor is used to generate a cryptographically secure PUF, which is LPN (learning parity with noise)-hard and thus cannot be broken by classical as well as quantum attacks. The PUF realizes a POK, which consists of a series of RO pairs. The frequency of each RO in a RO pair is counted, and a subtractor generates the difference between them. The difference is defined as the confidence value, which forms the entropy source of the RO. An FPGA implementation of thus PUF was provided in \citep{Jin-FPGA-trapdoor}. A TRNG generates the challenges for the PUF, and an LPN-hard POK produces the response.

\subsubsection{Quantum-secure Authentication of PUFs}
Apart from PUF architectures, the authentication protocols utilized to ensure secure communication in cryptographic IPs are also at risk against quantum adversaries. An example structure of general authentication protocols consisting of a PUF is shown in Figure~\ref{fig:PUF-EAP}. Any such authentication protocol consists of an enrollment phase where the general challenge-response pairs are generated and enrolled in the database. When the protocol is deployed in an IP or any other channel to provide security and verify a specific set of challenges, the verification phase starts. In order to ensure such protocols are secure against quantum attacks, new authentication protocols have been proposed keeping in mind the resilience against quantum attacks. A comprehensive summary of such protocols has been provided in the table \ref{tab:QA-protocol}. Many of these authentication protocols derive their security mostly from NP-hard problems like LPN and thus can be termed secure against quantum attacks.   
\vspace{0.5ex}

\noindent\textbf{PUF-HB}: In \citep{Hammouri-PUF-HB}, a protocol is proposed that uses an amalgamation of PUF and the Hopper Blum (HB) function (thus, it is dubbed PUF-HB). Authentication protocols using HB function were first proposed by Hopper and Blum \citep{Hopper-Blum} and have major advantages like the reduction of power consumption for pervasive networks. The security of the protocol is based on the hardness of the LPN problem making the PUF resilient against classical as well as quantum attacks. The PUF-HB authentication protocol consists of two entities: the tag and the reader. The tag uses the PUF response and a few bit vectors to compute a response for a given challenge and sends it to the reader. The reader computes the same using the database and authenticates the response as correct or incorrect for a series of authentication rounds. After the predefined number of rounds is completed, the reader authenticates the response only when the number of incorrect responses is within a predetermined level of tolerance. Given the fact that the LPN problem is noise parity (NP) hard, the PUF-HB protocol is provably secure against active attackers. The authors claim that the given authentication protocol does not store the intermediate PUF response, which is used to compute the function. Hence, there is no way that the attacker can tamper the PUF to extract the PUF response and attempt to clone the PUF-HB protocol, which proves the PUF to be tamper-resilient as well.

\vspace{0.5ex}

\noindent\textbf{QR-PUF authentication}: Quantum PUFs cannot be cloned, and this is a major advantage of quantum PUFs over classical PUFs. In \citep{Giulio-PRA}, the authors extend the metrics of uniqueness and robustness of a classical PUF to characterize quantum PUFs like QR-PUFs \citep{Skoric-2012}. Here, the authors design a generalized authentication scheme that can be applied to both classical and quantum PUFs like QR-PUFs (described in subsection \ref{subsec:pufs-defense}). By comparing the quantum PUFs against classical PUFs, the authors conclude that classical measurements can be observed without disturbing the current state of a classical PUF; thus, the response can be easily copied. Thus, for a classical PUF after $q$ interactions, an adversary will know exactly $q$ CRPs, which can be used to create a clone of the classical PUF. But for quantum PUFs, it is difficult to copy quantum states, and also measurements will introduce errors in the quantum response. Thus after $q$ interactions, the adversary will have less than $q$ CRPs to clone the quantum PUF. This makes a quantum PUF like QR-PUF much superior in terms of unclonability. The authors also mention that quantum challenges are non-orthogonal in nature; thus, a quantum PUF cannot distinguish between states. This non-orthogonality of challenges helps the no-cloning property but introduces errors in measurement, thus undermining robustness. 

\vspace{0.5ex}
\begin{table*}[t]
\caption{Summary of quantum authentication protocols}
\scriptsize
\begin{tabular}{|l|l|l|l|l|}
\hline
\begin{tabular}[c]{@{}l@{}}Authentication\\ scheme\end{tabular} & Type of PUF used & Algorithm used & Advantages/Contribution & Limitations \\ \hline
\begin{tabular}[c]{@{}l@{}}PUF-HB\\ \citep{Hammouri-PUF-HB}\end{tabular} & Classical PUF & \begin{tabular}[c]{@{}l@{}}Hopper Blum\\ function\end{tabular} & \begin{tabular}[c]{@{}l@{}}Tamper resilient, unclonable\\ LPN-hard, secure against \\ classical and quantum\\ attacks\end{tabular} & \begin{tabular}[c]{@{}l@{}}Performance against \\ quantum attacks like \\ challenge estimation, \\ quantum emulation \\ not shown\end{tabular} \\ \hline
\begin{tabular}[c]{@{}l@{}}QR-PUF authentication\\ \citep{Giulio-PRA}\end{tabular} & QR-PUFs & \begin{tabular}[c]{@{}l@{}}Entity \\ authentication\\ protocols\end{tabular} & \begin{tabular}[c]{@{}l@{}}Characterization of \\ QR PUF metrics like \\ uniqueness and reliability\\ against classical PUFs\end{tabular} & \begin{tabular}[c]{@{}l@{}}No suggestion \\ provided to improve\\ reliability of QR-PUFs\end{tabular} \\ \hline
\begin{tabular}[c]{@{}l@{}}Quantum-secure \\ authentication\\ \citep{Goorden-Optica}\end{tabular} & \begin{tabular}[c]{@{}l@{}}Classical optical\\ PUF\end{tabular} & \begin{tabular}[c]{@{}l@{}}Application\\ of quantum\\ challenges to \\ optical PUF\end{tabular} & \begin{tabular}[c]{@{}l@{}}Resilient against emulation \\ attacks\end{tabular} & \begin{tabular}[c]{@{}l@{}}Metrics like reliability\\ and uniqueness of the \\ PUF not discussed\end{tabular} \\ \hline
\begin{tabular}[c]{@{}l@{}}Continuous-variable \\ authentication\\ \citep{Nikolopoulos-SCReports}\\ \citep{Nikolopoulos-PhysRevA}\end{tabular} & \begin{tabular}[c]{@{}l@{}}Classical optical\\ PUF\end{tabular} & \begin{tabular}[c]{@{}l@{}}Application\\ of quantum\\ challenges using\\ coherent states\\ of light\end{tabular} & \begin{tabular}[c]{@{}l@{}}Resilient against emulation \\ attacks\end{tabular} & \begin{tabular}[c]{@{}l@{}}Metrics like reliability\\ and uniqueness of the \\ PUF not discussed\end{tabular} \\ \hline
\end{tabular}
\label{tab:QA-protocol}
\end{table*}

\noindent\textbf{Quantum-secure authentication (QSA)}: The QSA technique proposed in \citep{Goorden-Optica} uses a quantum state as a challenge instead of classical states. The authors argue that the traditional optical physically unclonable key (PUK) based authentication protocols rely on classical challenges. Thus, if the attacker has access to the authentication protocol and possesses sufficient information about the CRP characteristics, then they can intercept the classical challenges sent to the verifier. The adversary can then emulate an attack by measuring such challenges and sending them to the verifier and can easily receive correct responses. In that case, the attacker does not need to clone the quantum PUF. This kind of attack is termed as an "emulation attack''. However, if the challenges are based on quantum states which are inherently unclonable, then the attacker will not be able to clone the challenges and obtain a valid response. The authentication protocol authenticates a classical multiple-scattering key by interrogating it with a light pulse and verifying the spacial structure of the reflected light.

\vspace{0.5ex}

\noindent\textbf{Continuous variable authentication protocol}: In \citep{Nikolopoulos-SCReports}, the authors propose an authentication protocol using optical PUFs. The major proposition is the usage of quantum challenges instead of classical ones, similar to the QSA technique described above. The proposed authentication scheme aims at using quantum challenges derived from random coherent states of light. The quadrature structures formed by the scattering of light originating from the laser source is analyzed by a coarse-grained homodyne detector. The homodyne detection helps in extracting information from the phase and frequency of the quadrature light structures. In \citep{Nikolopoulos-PhysRevA}, the security of the above authentication protocol has been analyzed against emulation attacks. The authors use Helevo's bounds and Fano's inequality to prove that the continuous variable scheme is secured against emulation attacks.

The concept of security against quantum attacks has emerged due to advancements in the fields of the quantum supercomputer. The impact of such discoveries on hardware security is still an open-ended problem, and a lot needs to be investigated to assure security for PUFs in the quantum world that can solve active quantum attacks. Quantum superiority seems to be a double-edged sword, which has provided many advancements in security primitives like the design of quantum PUFs. The inherent principles of unclonability further strengthen the PUF performance and increase tamper resilience and man-in-middle attacks for quantum PUFs. On the other hand, the presence of quantum technology introduces several vulnerabilities in existing security primitives, compromising several cryptographic and security protocols. Quantum technology profits the defended as well as the attacker. In our next section, we try to look at several attack strategies against PUFs in the post-quantum era.

\subsection{Attacks against PUFs in Post Quantum Era}

Attacks can occur against a) classical PUFs or b) quantum PUFs. Attack strategies against classical PUFs are a well-researched topic and include a wide range of attacks, as discussed below. Attacks against quantum PUFs is a relatively new topic and still needs a lot of work. 

\subsubsection{Attacks against Classical PUFs}
In general, the attacks against classical PUFs can be divided into two major subsections a) classical attacks where the adversary does not possess any quantum processing capability and b) quantum attacks where the attacker has quantum computation competence, and we discuss the state-of-art works in both the sections below.

Attacks against classical PUFs can be divided into three main categories, namely a) invasive attacks, b) semi-invasive or side-channel attacks, and c) non-invasive or software/ machine learning (ML) attacks. Invasive attacks on PUFs are capable of altering the physical structure of the PUFs leading to the extraction of decrypted data and completely changing the PUF configuration. An example attack can be seen in \citep{Helfmeier-CCS-13}, where the attackers permanently modify the device security fuses by invasive focused-ion-beam (FIB) circuit edit and decipher sensitive data using backside micro probing. Semi-invasive attacks on PUFs include an amalgamation of fault-injection/side-channel attacks along with an ML framework to decode sensitive CRP behavior from the PUFs. In \citep{Tajik-FTDC}, the attackers use laser fault injection attacks coupled with ML algorithms to alter configurations and compromise the security of XOR-arbiter and RO PUFs implemented in programmable logic devices (180nm). For non-invasive or ML attacks, the attacker uses a small subset of the CRPs of the PUF and tries to build a model representing the challenge-response behavior of the PUF. The first ML attack against PUFs was implemented in \citep{Gassend-ML}, and after that, different types of ML attacks have been developed against PUFs. ML attacks against classical PUFs can be divided into two types a) empirical attacks and b) PAC (probably approximately correct) framework. Empirical ML approaches \citep{Ruhrmair-CCS} do not have any predefined level of confidence or accuracy, whereas the PAC framework develops models with specified levels of accuracy, as seen in papers like \citep{Ganji-PAC,ganji2015attackers}.

\subsubsection{Attacks against Quantum PUFs}
The security of quantum readout PUFs against challenge estimation attacks has been shown in \citep{Skoric-JQI}. This is an example of a classical attack possible against quantum PUFs, and the authors provide an analysis of the security of the respective PUF against such attacks. In this scenario, the adversary has access to the verification protocol and intercepts the challenge quantum state. The attacker then performs a measurement on the challenge state and reconstructs the challenge to emulate the PUF.  Though this attack can be implemented against quantum secure authentication protocol, which uses a classical challenge, it cannot be used against protocols that use quantum states as challenges, as in this case, the attacker cannot measure a quantum state, making the attack ineffective. The QSA \citep{Goorden-Optica} and continuous variable authentication protocols \citep{Nikolopoulos-SCReports} described above are robust against such forms of attack. The QR PUF has also been analyzed to be resilient against such attacks, as mentioned in \citep{Skoric-JQI}. The authors claim that the number of photons detected in the verification protocol is lower by more than a factor of the quantum security parameter ($S$) \footnote{$S=\frac{K}{n}$, where $K$ is the order of the information contained in the challenge and n is the total number of photons} defined in the paper making the QR PUF robust against such classical attack.  

Another form of attack is based on quantum interactions, where the attacker has quantum capabilities and the system which is attacked consists of a quantum PUF (QPUF). One example of such an attack is the quantum emulation attack, as described in \citep{arapinis2019quantum}, where the adversary uses quantum emulation (QE) algorithm \citep{marvian2016universal} to attack a QPUF. The authors mention QPUFs as quantum channels with the standard inherent quantum qualities of unclonability, robustness, and collision resistance. The authors further explain that due to the property of collision resistance of the QPUFs, they can be modeled by unitary transformations, and thus such PUFs are termed in the paper as unitary quantum PUFs or UQPUFs. The authors prove that quantum emulation attacks can easily compromise the unclonability property of the UQPUFs. If the adversary can choose the target challenge themselves, then quantum primitives are unable to provide security against such adversaries. Also, if the attacker has access to the challenge and can interrogate the primitive after receiving the challenge, then no current quantum primitive can protect against such attacks. In short, the authors show that the unclonability property of UQPUFs can be compromised by quantum emulation attacks. Another important aspect of emulation attacks has been shown in \citep{Fladung-2019}. Here the authors use a similar framework of intercept-emulation attacks investigated in \citep{Nikolopoulos-PhysRevA}; they assume that the adversary has access to a copy of the numerical CRPs and intercepts the measurement apparatus to make an educated guess regarding the other security parameters required to induce the correct response state of the verifier. The paper discusses the robustness of continuous variable authentication protocol against three types of emulation attacks and discusses the security metrics, which ensures the security of the protocol against all the mentioned attacks.

The metric which signifies the robustness of classical PUFs against modeling attacks is unclonability. While, it has been stated in several papers like \citep{Giulio-PRA}, \citep{Skoric-2012}, that quantum PUFs are superior to classical PUFs in terms of unclonability. But with the advent of novel attacks like quantum emulation and classical challenge estimation attacks as described before, the unclonability property of quantum PUFs can be compromised. With the advent of new attack strategies, new authentication protocols have also been proposed which claim to protect against the above attacks, but all these protocols have their common set of advantages and limitations, as described in table \ref{tab:QA-protocol}. Also, the property of unclonability adversely affects other metrics of quantum PUFs like reliability. Thus, an optimum trade-off of the above metrics is required that can produce an unclonable, as well as a reliable quantum PUF. 


 
}
\section{General Conclusion and Lessons Learned}\label{sec:conclusion}
\sloppypar{
Albeit being broadly acknowledged and studied, applications of quantum technology in cryptography, and in particular, physical security, need more effort and more cooperative attention to become widely-deployed, global solutions. 
In this context, one of the key challenges facing the researchers is the realization of a complex system composed of various classical and quantum modules. 
For such a system, a solid foundation must be laid for the assessment of security, and accordingly, for the definition of metrics and design of tests matching the requirements of real-life applications. 
This, of course, demands standardization efforts, supported and promoted by national and international standardization bodies. 
An essential part of these standardization activities is examining whether new proposals for post-quantum cryptosystems can stand
the test of time, and more specifically, investigating the vulnerability of these systems to various attacks. 

With respect to this observation, fundamental studies on the resistance of post-quantum cryptosystems to quantum-enhanced and classical, physical attacks should be pursued. 
In order to complement these studies, it is necessary to come up with approaches concerning how to harness the power of quantum mechanics in physical primitives that can prevent physical attacks. 
These primitives must be in compliance with real-world conditions, e.g., capabilities of users and attackers, to provide the required security, when the current physical countermeasures are no longer effective. 
By conducting an extensive review of the literature, this survey attempts to provide useful insights into the challenges associated with these matters, summarized as follows. 

\begin{itemize}[leftmargin=0.25cm]

\item \textbf{Side-channel analysis:}
Along with the development of numerous quantum-resistant cryptographic schemes, as discussed above, their standardization has been in progress. 
While those post-quantum algorithms may be proven quantum-resistant, they are not necessarily resistant against classical side-channel attacks. 
These attacks can be extremely powerful, as they exploit physical data leakage from the device. Side-channel attacks on post-quantum algorithms can be still thought to be less harmful, as they are not implemented into commercial systems yet. 

To protect post-quantum cryptographic schemes against power analysis, first-order masking and hiding are used to introduce the noise; however, both have been broken. Hiding is done by inserting dummy and by shuffling operations, such as randomization of row iteration for matrix-vector multiplication. Timing attacks are halted by the introduction of constant operations, such as constant multiplication, or constant calling of hash functions. Countermeasure, which would secure against timing and SPA attacks, proposed in~\citep{ref30}, requires redundant addition always being performed, making runtime independent of secret value and message. Proposed countermeasures against fault attacks which exploit vulnerabilities in error correction, require encryption and decryption to be implemented as a nested procedure inside the scheme, because if the correct decryption of the message means message does not have any errors, as proposed in~\citep{ref33}. For fault analysis attacks which introduce randomization, skipping and zeroing, simple checking and comparisons of the correctness of secret key are proposed to prevent the attacks.

Compared to SCA launched on classical devices, the attacks on post-quantum cryptographic schemes do not have to break strong countermeasures as the first step. Their development is in very early stages, compared to countermeasures against SCA on classical algorithms, such as sense-amplified-based logic (SABL), wave dynamic differential logic, and t-private logic circuits. Most of the post-quantum countermeasures are not provably secure, and they protect against a specific attack while being compromised by others. 

However, the current security of post-quantum algorithms lies in algorithm complexity. These algorithms have computations almost inherent to secret keys through the various matrix and linear algebraic properties. The attacker is challenged to find computations dependent on secret keys and values, which can be controlled, in order to exploit the secret key. 
Further research must be done in assessing the vulnerabilities of post-quantum algorithms and developing countermeasures that would protect against various side-channel attacks, as well as invasive physical probing, which has not been evaluated yet.

Side-channel attacks in the quantum world are also considered to take advantage of the quantum device itself. Data acquisition does not benefit from quantum speed-ups, so the quantum algorithm application is not evaluated for this application. However, the key search is the SCA step, which can greatly benefit from quantum speed-up, and Grover's quantum algorithm seems to be the best option for this purpose, regardless of the limitation of requiring unstructured input data.  

The danger of SCA on post-quantum cryptographic schemes that are heavily developed and evaluated is large. Due to the novelty of many post-quantum cryptographic schemes, many side-channel vulnerabilities are not yet evaluated. Current countermeasures for extensively researched post-quantum algorithms have an ad-hoc design, which protects against a specific version of SCA, while it does not protect against other attacks. Security of most novel post-quantum cryptographic schemes relies on original algorithm structure, and lack of SCA attempts. The use of quantum algorithms may not be useful for SCA data acquisition from classical schemes. However, its usefulness could be evaluated for the search of post-quantum computation intermediate values, which depend on the secret key and controlled message.

\vspace{0.5ex}    
\item \textbf{Random number generators:} Throughout the journey of random number generators from Vernam's one-time pad to Internet-of-Things (IoT) devices, the ability to communicate securely and compute efficiently remains a common goal. 
In particular, for the IoT technology, generators with smaller implementation overhead, being cheap enough, and fast enough for widespread use are preferred. 
QRNGs can be thought of as promising candidates fulfilling these needs, under the premise that they can circumvent practical difficulties with TRNGs, namely limited generation rate, no guarantee on intrinsic randomness, challenges with failure and online tests. 
The latter problem is indeed not specific to TRNGs since QRNGs also require quantum sources to be examined frequently in order to avoid any defect and interference. 
Hence, it is essential to meet the demand for highly accurate and reliable testing approaches, especially self-testing methods. 
Fortunately enough, these methods have been studied in the context of TRNGs, and interestingly, proposals have been made to be adopted by the TRNGs and QRNGs simultaneously. 
Regardless of how a self-testing is built, entropy estimation is a key building block of that. 

Another area of research related to this is randomness amplification that has been reviewed in this survey. 
More specifically, if the results of a test at the startup stage demonstrate a lack of enough entropy that can be extracted from the QRNG, randomness amplification techniques come into play. 
As explained before, these techniques, however, may suffer from several problems in practice. 
It is particularly true concerning how confidently the true randomness of a source can be determined. 
We believe that this can be seen as a future research direction.  
Furthermore, it is of great importance to come up with high-quality, electronic chip-based quantum sources that can be easily embedded in commercial devices. 
For such devices, not only the generation rate, but also the implementation cost and the compatibility with off-the-shelf systems are crucial aspects of the design. 

To sum up, we stress that advances in the theory and implementation of QRNGs will eventually lead to the design of more robust and practically feasible encryption, key distribution, and in general, cryptography and other applications relying on true random numbers. 

\vspace{0.5ex}
\item \textbf{Physically unclonable functions:} Silicon PUFs are defined as one of the major developments of hardware security primitives and entail a mature research field. 
Compared to classical PUFs, post-quantum research in PUFs is a nascent field. In this survey, we presented a comprehensive overview of the current research that has been conducted so far related to PUFs in the post-quantum era. We have divided our study into two major fields; one concerns with the defense and improvement of PUF architectures to protect against new innovations concerning quantum adversaries. Here, we have discussed PUF architectures that originate from quantum phenomena along with their advantages and disadvantages.
    
We have also discussed the different authentication protocols involving PUFs that have been proposed to defend against quantum-adversaries along with  a thorough survey of classical PUFs that are claimed secure against existing quantum attacks. The other section involved different attack strategies that have emerged with the quantum advancement. To provide perspective, we outlined the major attack strategies against classical PUFs and then moved on to different forms of attacks developed so far both against quantum PUFs. To the best of our knowledge, this is one of the preliminary surveys which aims at reviewing the state-of-art research in post-quantum PUFs. 
    
In retrospect, the research performed so far in the post-quantum PUF era includes both defense and attack mechanisms. Quantum phenomena have been used to generate useful unclonable PUF architectures. Authentication protocols that are robust against quantum attacks have been proposed. Yet, there still exists a lack of organization, and the work done so far can be termed ad-hoc. Quantum computing is a powerful weapon that can impact the field of security as a double-edged sword. The advantages of unclonability in quantum domain strengthens PUF architectures against quantum attackers but also undermines the reliability metric of the PUF. Thus, a prospective field of research may include the improvement of reliability for quantum PUFs allowing an optimum unclonability. The existence of NP-hard authentication protocols provides potential prospective security solutions against quantum attacks. Nevertheless, better evaluation algorithms are required, and a new set of metrics needs to be formulated to ensure the security PUFs against quantum attacks. Once a proper set of metrics and evaluation tests are developed to evaluate existing PUF architectures against quantum attacks, robustness against quantum attackers can be verified.   
\end{itemize}
}

\section{Acknowledgments}
The author would like to acknowledge the support of AFOSR under award number FA 9550-14-1-0351. 

\label{sect:bib}
\bibliographystyle{spbasic.bst}
\small
\bibliography{references}

\appendix


\end{document}